\documentclass[journal,twocolumn,10pt,twoside]{IEEEtran}
\normalsize

\usepackage{cite}

\ifCLASSINFOpdf
\else
\fi

\usepackage{url}
\usepackage{amsmath}
\usepackage{amssymb}
\usepackage{cite}
\usepackage{graphicx}
\usepackage{algorithm}
\usepackage{algpseudocode}
\usepackage{subfigure}
\usepackage{multirow} 
\usepackage{longtable}
\usepackage{threeparttable}
\usepackage{caption3}
\usepackage{array}
\usepackage{cases}
\usepackage{color}
\usepackage[T1]{fontenc}
\usepackage[utf8]{inputenc}
\usepackage{authblk}
\usepackage{subfigmat}

\usepackage{caption}
\usepackage{multirow}

\usepackage{float}

\usepackage{multirow, makecell}
\usepackage{booktabs}
\usepackage{epstopdf}
\usepackage{graphicx, subfigure}

\newcommand{\rf}{\mathrm{rf}}
\newcommand{\dc}{\mathrm{dc}}

\newtheorem{remark}{Remark}

\begin{document}

\title{Signal and System Design for Wireless Power Transfer : Prototype, Experiment and Validation}

\author{Junghoon~Kim,
        Bruno~Clerckx,
        and~Paul~D.~Mitcheson
        
\thanks{The authors are with the Department of Electrical and Electronic Engineering, Imperial College London, London SW7 2AZ, U.K. (e-mail: junghoon.kim15, b.clerckx, paul.mitcheson@imperial.ac.uk). This paper is an expanded version from the IEEE MTT-S Wireless Power Transfer Conference, Taipei, Taiwan, May 10-12, 2017. This work has been partially supported by the EPSRC of UK, under grant EP/P003885/1.}%
}

\maketitle

\begin{abstract}
A new line of research on communications and signals design for Wireless Power Transfer (WPT) has recently emerged in the communication literature. Promising signal strategies to maximize the power transfer efficiency of WPT rely on (energy) beamforming, waveform, modulation and transmit diversity, and a combination thereof. To a great extent, the study of those strategies has so far been limited to theoretical performance analysis. In this paper, we study the real over-the-air performance of all the aforementioned signal strategies for WPT. To that end, we have designed, prototyped and experimented an innovative radiative WPT architecture based on Software-Defined Radio (SDR) that can operate in open-loop and closed-loop (with channel acquisition at the transmitter) modes. The prototype consists of three important blocks, namely the channel estimator, the signal generator, and the energy harvester. The experiments have been conducted in a variety of deployments, including frequency flat and frequency selective channels, under static and mobility conditions. Experiments highlight that a channel-adaptive WPT architecture based on joint beamforming and waveform design offers significant performance improvements in harvested DC power over conventional single-antenna/multi-antenna continuous wave systems. The experimental results fully validate the observations predicted from the theoretical signal designs and confirm the crucial and beneficial role played by the energy harvester nonlinearity.
\end{abstract}


%
\IEEEpeerreviewmaketitle

\section{Introduction}

\IEEEPARstart{I}{nterests} in radiative (far-field) Wireless Power Transfer (WPT) have been growing recently because WPT is an invaluable technology to energize a large number of low-power autonomous devices. It is viewed as an enabler for many emerging applications such as the Internet of Things, Wireless Sensor Networks, and for innovative wireless networks where radiowaves are used for the dual purpose of communicating and energizing \cite{Clerckx2019}. 
In particular, radiative WPT is attractive since it enables long-distance power delivery and small receiver form factors, in comparison with other technologies. 
A crucial challenge of radiative WPT system design is to maximize the harvested DC power subject to a transmit power constraint, or equivalently to enhance the end-to-end power transfer efficiency. 
To that end, the traditional line of research in the RF literature has been devoted to the design of efficient rectennas so as to increase the RF-to-DC conversion efficiency. 
It is well-established that a variety of technologies (e.g. CMOS, Diode) and topologies (e.g. single shunt, voltage multiplier) can be considered for rectenna designs \cite{Pinuela2013,Hagerty2004,Valenta2014}. 

\par Aside rectenna design, a new and complementary line of research on communications and signal design for WPT has emerged recently in the communication and signal processing literature \cite{Zeng2017}.
Indeed the amount of DC power that can be harvested is not only a function of the rectenna design but also of the transmit signal strategy and of the wireless propagation channel condition. 
In other words, the transmit signal design has a major impact on the end-to-end power transfer efficiency as it influences the signal strength at the input of the rectenna but also the RF-to-DC conversion efficiency of the rectifier. 
Four different kinds of transmit signal design strategies have been proposed, specifically for WPT purposes, to boost the received DC power.
\par
A \textit{first} strategy is to design (energy) \textit{waveforms} in order to exploit the rectenna nonlinearity and boost the RF-to-DC conversion efficiency $e_{\rf-\dc}$. 
Previous studies have observed that multisine waveforms can increase $e_{\rf-\dc}$ \cite{Trotter2009},\cite{Boaventura2011}, and that high Peak-to-Average Power Ratio (PAPR) waveforms enhance $e_{\rf-\dc}$ \cite{Collado2014}.
A systematic waveform design methodology for WPT was first proposed in \cite{Clerckx2016}.
Waveforms can be designed with and without Channel State Information at the Transmitter (CSIT) depending on the frequency selectivity of the channel.  
The optimal design of channel-adaptive waveform in \cite{Clerckx2016} results from a tradeoff between exploiting the channel frequency selectivity (so as to maximize the RF-to-RF transmission efficiency $e_{\rf-\rf}$) and the energy harvester (EH) nonlinearity (so as to boost the RF-to-DC conversion efficiency $e_{\rf-\dc}$).
Due to the rectifier nonlinearity, the optimal waveform design can be obtained as the solution of a non-convex optimization problem, which is not easily implemented in a real-time system.
Strategies for reducing waveform design complexity have therefore been introduced in \cite{Clerckx2017,Huang2017,Moghadam2017}. Moreover, since CSI needs to be acquired to the transmitter, a proper joint design of the waveform and the channel acquisition strategy needs to be considered, and a possible solution is to design the waveform for WPT based on a limited number of feedback bits \cite{Huang2018}.
\par 
A \textit{second} strategy is to design multi-antenna (energy) \textit{beamforming} in order to increase the RF input power of the energy harvester and therefore enhance the RF-to-RF transmission efficiency $e_{\rf-\rf}$.
This strategy also requires an appropriate CSIT acquisition scheme for WPT \cite{Xu2014}.
Similarly to wireless communications, the simplest form of beamforming is Maximum Ratio Transmission (MRT)\cite{Lo1999}.
Alternative techniques to multi-antenna beamforming, also enabling directional/energy focusing transmission, consist in retrodirective and time-modulated arrays \cite{Masotti2016} and time-reversal techniques \cite{Ku2017}.
Waveform and multi-antenna beamforming can be combined so as to optimally exploit the beamforming gain, the channel frequency diversity gain and the nonlinearity of the rectifier \cite{Clerckx2016,Huang2017}.

\par
A \textit{third} strategy is to design (energy) \textit{modulation} for single-carrier transmission.
In contrast to the energy waveform that commonly relies on an optimized deterministic multisine/multi-carrier, the energy modulation induces random fluctuations of a single-carrier. Similarly to the energy waveform, the design of the energy modulation aims at exploiting the nonlinearity of the rectifier to boost the RF-to-DC conversion efficiency $e_{\rf-\dc}$. Indeed, as a consequence of the energy harvester nonlinearity, the RF-to-DC conversion efficiency $e_{\rf-\dc}$ differs depending on whether the rectifier input signal is modulated or not \cite{Clerckx2018a}.
For instance, a real Gaussian modulation offers a higher harvested DC power than a circularly symmetric complex Gaussian modulation \cite{Varasteh2017}. 
A new modulation scheme based on flash signaling (a form of on-off keying distribution with low probability of high amplitude signals) has recently been introduced in \cite{Varasteh2018}. It exploits the rectifier nonlinearity by transmitting signals of very high amplitudes with low probability. Flash signaling was shown to outperform a real Gaussian modulation and maximize the amount of harvested DC power. Flash signaling-based energy modulation can also be combined with multi-antenna so as to additionally exploit a beamforming gain.

\par
A \textit{fourth} strategy is to use phase sweeping \textit{transmit diversity} in a multi-antenna WPT setup to boost the RF-to-DC conversion efficiency \cite{Clerckx2018}. Transmit diversity aims at artificially inducing fast fluctuations of the wireless channel at the input of the rectifier using dumb transmit antennas. Those fluctuations are shown to improve the RF-to-DC conversion efficiency thanks to the rectifier nonlinearity.
Interestingly, transmit diversity does not rely on CSIT and demonstrates that
multiple transmit antennas can be beneficial to WPT even in the
absence of CSIT. 
\par 
The theoretical performance benefits of the aforementioned four signal strategies have been studied in the literature, based on simplified linear and nonlinear energy harvester models. Since the theoretical analysis relies on numerous assumptions, commonly made to simplify the signal and system design, it remains to be seen whether those emerging signal designs for WPT still deliver the expected benefits in a realistic setup.  In particular, aside the crucial nonlinearity and nonidealities of the energy harvester, real-world experimentation of WPT is subject to numerous impairments such as amplifier nonlinearity and gain/phase offset, that are neglected, and can be overlooked, in any theoretical analysis. This calls for prototyping and experimenting those emerging signal strategies to assess their real-world performance and validate the feasibility of the underlying signal theory for WPT.

\par 
There have been studies on WPT prototyping, in both the RF and the communication literature. In the RF literature, multisine waveforms have been experimented in \cite{Trotter2009,Valenta2015,valenta_durgin_2016} and the corresponding $e_{\rf-\dc}$ has been measured. These experiments were performed using open-loop based prototypes with static and heuristic waveforms fed directly into the rectifier, not using closed-loop based architecture with channel-adaptive (relying on CSIT so as adjust the transmission strategy dynamically as a function of the wireless frequency-selective propagation channel) and optimized waveforms transmitted over-the-air. In the communication literature, emphasis has been put on closed-loop based adaptive beamforming with a multi-antenna transmitter, as shown in e.g. \cite{Choi2017,Choi2018,Yedavalli2017,Claessens2018,Abeywickrama2018}. These works studied channel acquisition techniques, and over-the-air feedback, for WPT and focused on increasing $e_{\rf-\rf}$ through adaptive beamforming. 
\par Recall nevertheless that maximizing the end-to-end power transfer efficiency is not achieved by maximizing $e_{\rf-\rf}$ and $e_{\rf-\dc}$ independently \cite{Zeng2017,Clerckx2018b}. This is because $e_{\rf-\rf}$ and $e_{\rf-\dc}$ are coupled due to the rectifier nonlinearity. This calls for systematic signal strategies that maximize the overall power transfer efficiency $e_{\rf-\rf} \times e_{\rf-\dc}$ by jointly accounting for the effect of the wireless channel and the harvester nonlinearity \cite{Zeng2017,Clerckx2016,Clerckx2018b}, and therefore completely bridge the RF and communication approaches. The first prototype to demonstrate the feasibility and over-the-air performance of waveform strategies that are adaptive to the wireless channel, account for the harvester nonlinearity and maximize $e_{\rf-\rf} \times e_{\rf-\dc}$ appeared in \cite{Kim2017}. 

\par In this paper, we build upon the prototype of \cite{Kim2017}, and implement all the four recently developed signal design strategies, namely waveform, beamforming, modulation, transmit diversity. The performance gain and feasibility of all those four strategies, and combination thereof, in real-world environments is assessed and verified experimentally for the first time\footnote{Recall that \cite{Choi2017}, \cite{Choi2018},\cite{Yedavalli2017},\cite{Claessens2018}, and \cite{Abeywickrama2018} focus on beamforming-only techniques where beamforming is optimized/designed for WPT. None of them focuses on energy modulation (designed to maximize the harvested DC power), waveform (designed to maximize harvested DC power), or transmit diversity. Any modulation or waveform used in those papers are conventional communication modulation/waveform, not signals designed/optimized for WPT.} in the literature. In particular, we ask ourselves the following questions: Can we establish an experimental environment of open-loop and closed-loop WPT and verify the advantages of systematic signal design for WPT (including waveform, beamforming, modulation, transmit diversity) through experimentation? Can we confirm theory from measurements? Can we validate or invalidate the linear and nonlinear energy harvester models used in the WPT and Wireless Information and Power Transfer (WIPT) literature? The contributions of the paper are summarized as follows. 
\par \textit{First}, we design, prototype and experiment a WPT system that can operate in both open-loop and closed-loop modes. The setup consists of three important blocks, namely the channel acquisition, the signal optimization and generation, and the energy harvester(s). Software Defined Radio (SDR) is used to implement a wireless power transmitter and a channel estimator, and various rectennas with single-diode and voltage doubler rectifiers are designed to work as energy harvesters. Leveraging the flexibility and reconfigurability of the SDR, it is possible to implement various transmission signal design and CSI acquisition strategies in one set of experimental equipment. In its open-loop WPT mode, the architecture does not rely on any CSIT (and therefore the channel acquisition module), though still increases the harvested DC power by using energy modulation and transmit diversity. In its closed-loop WPT mode, the architecture relies on a frame structure switching between a channel estimation/acquisition phase and wireless power transmission phase. Channel acquisition is performed every second and transmit signal is generated according to a joint waveform and beamforming design to maximize $e_{\rf-\rf} \times e_{\rf-\dc}$. 
\par
\textit{Second}, we implement the four signal design strategies mentioned above, namely waveform, beamforming, transmit diversity and energy modulation, in the prototype. The real over-the-air performance are assessed experimentally for each of the strategies and for a combination thereof. This contrasts with other WPT prototyping works that focus on the adaptive beamforming approach only, e.g. \cite{Choi2017,Choi2018,Yedavalli2017,Claessens2018,Abeywickrama2018}, or on testing conventional/non-adaptive (not WPT-optimized) waveform \cite{Trotter2009,Valenta2015,valenta_durgin_2016,Collado2014}.

\par
\textit{Third}, the performance (in terms of harvested DC power) of the WPT architecture is investigated in a variety of deployments, including frequency flat (FF) and frequency selective (FS) channels, and under static and mobility conditions. Experiments highlight the suitability of each signal design under various propagation conditions and the role played by various parameters such as the channel frequency selectivity, the velocity, the number of tones, the number of transmit antennas, the signal bandwidth and the rectenna design.
\par 
\textit{Fourth}, and importantly, the experimental results of the various signal strategies confirm and validate the observation made from the theory of the rectifier nonlinearity and the signal designs proposed and developed in \cite{Clerckx2016,Clerckx2017,Clerckx2018,Clerckx2018a,Varasteh2017,Varasteh2018}. In particular, the following observations made from the theory are fully confirmed in the experiments: 1) The diode nonlinearity is fundamental and beneficial to WPT performance and is to be exploited in any systematic transmit signal design for WPT and WIPT; 2) The linear model of the EH, obtained by ignoring the rectifier nonlinearity, is not validated by experiments and measurements and leads to poor signal designs; 3) The wireless propagation channel and fading has a significant impact on WPT signal design and system performance; 4) A systematic WPT signal and system design has a big influence on the energy transfer efficiency with and without CSIT; 5) CSI acquisition and channel-adaptive waveforms are essential to boost the performance in frequency-selective channels; 6) Multiple antennas can be used in conjunction with transmit diversity to improve the energy transfer efficiency without CSIT; 7) Energy waveform and modulation can be used in conjunction with beamforming to maximize $e_{\rf-\rf} \times e_{\rf-\dc}$ and achieve a combined gain.

\textit{Organization}: Section \ref{sec:rect_model} introduces the system model and Section \ref{sec:waveform_sec} presents theoretical performance analysis. Section \ref{sec:prototype} introduces the prototype design. Section \ref{sec:experiment} offers all experimental results and observations and Section \ref{sec:conclusion} concludes the work and discusses future works.

\textit{Notations}: Bold letters stand for vectors or matrices whereas a symbol not in bold font represents a scalar. $\left|.\right|$ and $\left\| . \right\|$ refer to the absolute value of a scalar and the 2-norm of a vector. $\mathbb{E}\{ .\}$ refers to the averaging/expectation operator.

%
%
\section{The System and the Signal Models}\label{sec:rect_model}
We consider a Multiple Input-Single Output (MISO) WPT system based on the four signal design strategies mentioned in the introduction. The general system model, along with the mathematical model of each signal design technique, are presented in this section.
%

\subsection{MISO WPT System Model}
The transmitter is equipped with $M$ transmit antennas and uses $N$ subbands, while the receiver is equipped with a single antenna.
The transmit signal at time $t$ on antenna $m$ is written as
\begin{align}
\begin{split}
 x_{m}(t) &= \sum_{n=0}^{N-1}s_{n,m}(t)\mbox{cos}\left( 2\pi f_{n}t+\phi_{n,m}(t)\right) \\
              &= \Re\Bigg\{ \sum_{n=0}^{N-1}\omega_{n,m}(t)e^{j2\pi f_{n}t}\Bigg\}
\end{split}
\end{align}
with $\omega_{n,m}(t) = s_{n,m}(t)e^{j\phi_{n,m}(t)}$ where $s_{n,m}(t)$ and $\phi_{n,m}(t)$ refer to the amplitude and phase of the subband signal on frequeny $f_{n}$ and transmit antenna $m$ at time $t$. 
Quantities $\mathbf{S}$ and $\mathbf{\Phi}$ are $N\times M$ dimensional matrices of the amplitudes and phases of the sinewaves with their $\left(n,m\right)$ entry denoted as $s_{n,m}(t)$ and $\phi_{n,m}(t)$. 
The average transmit power constraint is given by $\sum_{m=1}^{M}\mathbb{E}\{ \left|x_{m}\right|^{2}\}=\frac{1}{2}\left\| \mathbf{S} \right\|_{F}^{2} \le P$. 
Vector-wise, the transmit signal vector $\mathbf{x}(t)$ can be written as
\begin{equation}
 \mathbf{x}(t) = \Re\Bigg\{ \sum_{n=0}^{N-1}\mathbf{w}_{n}e^{j2\pi f_{n}t}\Bigg\}
\end{equation}
where $\mathbf{w}_{n} = \left[ \omega_{n,1}(t)  \cdots  \omega_{n,M}(t) \right]^{T}$.
\par 
The transmit signal propagates through a multipath channel. We assume that the (frequency-domain) channel coefficient $h_{n,m}(t)$ changes at a rate slower than the transmission signal and that the channel is effectively stationary within a single time frame (i.e., we drop the time dependency of the channel coefficients).
The received signal from transmit antenna $m$ is written as
\begin{equation}\label{ymt}
 y_{m}(t) = \sum_{n=0}^{N-1}s_{n,m}(t)A_{n,m}\cos\left(2\pi f_{n}t+\psi_{n,m}(t) \right)
\end{equation}
where the amplitude and phase $A_{n,m}$ and $\psi_{n,m}$ are such that
\begin{equation}
 A_{n,m}e^{j\psi_{n,m}(t)}=A_{n,m}e^{j(\phi_{n,m}(t)+\bar{\psi}_{n,m})}=e^{j\phi_{n,m}(t)}h_{n,m}
\end{equation}
and the frequency response of the multipath channel is given by $h_{n,m} = A_{n,m}e^{j\bar{\psi}_{n,m}}$ .
The channel vector $\mathbf{h}_{n}$ can be written as $\mathbf{h}_{n} = \left[ h_{n,1}  \cdots h_{n,M}  \right]$.

\par The total received signal is the sum of \eqref{ymt} over all transmit antennas, namely
\begin{align}
\begin{split}
 y(t) &= \sum_{m=1}^{M}\sum_{n=0}^{N-1}s_{n,m}(t)A_{n,m}\cos\left(2\pi f_{n}t+\psi_{n,m}(t) \right)  \\
       &= \Re\Bigg\{ \sum_{n=0}^{N-1}\mathbf{h}_{n}\mathbf{w}_{n}e^{j2\pi f_{n}t}\Bigg\}.
\end{split}
\end{align}
\par At the receiver, the signal $y(t)$ impinges on the receive antenna and is absorbed by the rectifier. A simple and tractable model of the rectenna, introduced in \cite{Clerckx2016}, is used in this paper for the analysis. The model expresses the output DC current as a function of the input signal $y(t)$ and relies on a Taylor expansion of the diode characteristic function. Following \cite{Clerckx2016}, the rectenna output DC power under perfect matching and ideal low pass filter is related to the quantity
\begin{equation}\label{diode_model_2}
 z_{\mathrm{DC}} = k_{2}R_{\mathrm{ant}}\mathbb{E}\{y(t)^{2}\} + k_{4}R_{\mathrm{ant}}^{2}\mathbb{E}\{y(t)^{4}\}
\end{equation}
with $R_{\mathrm{ant}}$ the antenna impedance and $k_{i} = \frac{i_{\mathrm{s}}}{i!(n v_{\mathrm{t}})^{i}}$ for $i=2,4$, where $i_{\mathrm{s}}$ is the reverse bias saturation current, $v_{\mathrm{t}}$ is the thermal voltage, $n$ is the ideality factor. The fourth order term $\mathbb{E}\{y(t)^{4}\}$ accounts for the rectifier nonlinearity.  As a reference, following \cite{Clerckx2016}, $k_2=0.0034$ and $k_4=0.3829$. Considering $R_{\mathrm{ant}}=50\Omega$, the coefficient of the fourth order term is 5630 times larger that the second order coefficient, and explains why nonlinearity is non-negligible.
%

\subsection{(Energy) Waveform and Beamforming}\label{subsec:waveform}
Various channel non-adaptive (not relying on CSIT) and adaptive (relying on CSIT) multisine waveform strategies for WPT have been proposed in the past few years and can be used in single-antenna as well as multi-antenna setup \cite{Clerckx2016,Clerckx2017}. Since those waveforms are deterministic, i.e. not modulated, we can drop the time dependency such that $\omega_{n}(t) = \omega_{n}$. 
\par
In Table \ref{wave_desc}, we highlight various waveform design methods and the mathematical representations of the waveform coefficients $\omega_{n}$ for single antenna and $\mathbf{w}_{n}$ for multi antenna system, assuming the CSI (in the form of the frequency-domain response $\mathbf{h}_{n}$ for all frequency component $n$) is available at the transmitter. All those waveforms can be expressed in closed-form and can therefore be implemented and tested in real-time over-the-air transmission. We do not consider the optimal waveform design of \cite{Clerckx2016,Huang2017,Moghadam2017} because they result from a convex/non-convex optimization problem that cannot be easily solved and implemented in real-time.

\subsection{(Energy) Modulation}\label{subsec:mod_design}
Energy modulation is another strategy for WPT to induce fluctuations of the transmit signal amplitude of a single carrier and boost the harvested DC power. In contrast to the multisine waveform that is deterministic, energy modulation carries information due to the randomness inherent from the modulation. However the modulation is designed in such a way that it maximizes the harvested DC power \cite{Varasteh2018}. In its simplest form, $M = 1, N = 1$, and the transmit signal $\omega_{n,m}(t)$ at time $t$ on carrier frequency $f_{0}$ can be written as 
\begin{equation}\label{modul_sig}
\omega_{n,m}(t) = \omega(t) = s(t)e^{j\phi(t)} 
\end{equation}
where $s(t) = \sqrt{2P}\sqrt{m_{I}(t)^{2} + m_{Q}(t)^{2}}$ and $\phi(t) = \tan^{-1} \Big ( \frac{m_{Q}(t)}{m_{I}(t)} \Big)$.
The message signal can be expressed in complex form as $m(t) = m_{I}(t) + jm_{Q}(t)$ and $m(t)$ is a normalized ($\mathbb{E}\{ \left|m(t)\right|^{2}\}=1$) complex baseband equivalent signal that represents the (energy) modulation symbol at time $t$. The coefficient $\sqrt{2P}$ is used to guarantee the average transmit power constraint $P$. 
%
%
\begin{table*}[t]
\centering
\caption{Various Waveform Design Methods and Descriptions}
\label{wave_desc}

\setlength{\tabcolsep}{0.7mm} 
\renewcommand{\arraystretch}{1.05} 

\begin{tabular}{|m{12mm}|m{12mm}|m{12mm}|m{47mm}|m{75mm}|m{10mm}|}

\toprule 
\hline Antennas & CSIT                    			 & Waveform Design Method		& Expression  & Description &Reference \\ \hline
\hline \multirow{5}{*}{SISO}	& no CSIT		& UP 	& $\omega_{n} = \frac{\sqrt{2P}}{\sqrt{N}}$  & The uniform power allocation (UP) simply assigns the same power to all frequencies components, with a zero phase. & \cite{Clerckx2016}  \\ \cline{2-6}

& CSIT		& ASS 	& $\omega_{n} = \left\{\begin{matrix} \sqrt{2P}e^{-j\bar{\psi}_{n}} & n=\bar{n} \\ 0 & n\neq \bar{n} . \end{matrix}\right.$  & The adaptive single sinewave (ASS) allocates power to the one frequency corresponding to the strongest channel $\bar{n}=\arg \max_{i}\left|h_{i}\right|$. This is the optimal solution for the linear EH model (2nd order term-only in \eqref{diode_model_2}), and therefore aims at maximizing $e_{\rf-\rf}$.  & \cite{Clerckx2016}  \\ \cline{2-6}

& CSIT		& UPMF 	& $\omega_{n} = \frac{\sqrt{2P}}{\sqrt{N}}e^{-j\bar{\psi}_{n}} $  & The uniform power allocation and matched fiter (UPMF) allocates the same amplitude for all frequencies components, but the channel phase is matched on each sinewave based on the CSIT.& \cite{Clerckx2016}  \\ \cline{2-6}
 
& CSIT		& MF		&  $\omega_{n} = A_{n}\sqrt{\frac{2P}{\sum_{n=0}^{N-1}A_{n}^{2}}} e^{-j\bar{\psi}_{n}}$  & The matched filter (MF) allocates power to all sinewaves proportionally to the frequency domain channel strengths. It is a particular case  of SMF with $\beta=1$.        & \cite{Clerckx2016} \\ \cline{2-6}
 
& CSIT		& MAX PAPR	& $\omega_{n} = \frac{1}{A_{n}} \sqrt{\frac{2P}{\sum_{n=0}^{N-1}\frac{1}{A_{n}^{2}}}} e^{-j\bar{\psi}_{n}}$   & The maximize PAPR (MAX PAPR) allocates power inversely proportional to the channel strength to maximize the PAPR at the rectifier input.      & \cite{Clerckx2016} \\ \cline{2-6}
 
& CSIT		& SMF	& $\omega_{n} = A_{n}^{\beta}\sqrt{\frac{2P}{\sum_{n=0}^{N-1}A_{n}^{2\beta}}} e^{-j\bar{\psi}_{n}}$   & The scaled matched filter (SMF) is a low-complexity multisine waveform design strategy motivated by observations of the optimized signal. $\beta$ is a scaling factor, whose choice results from a compromise between exploiting the EH nonlinearity and the channel frequency selectivity, and therefore aims at maximizing $e_{\rf-\rf} \times e_{\rf-\dc}$. & \cite{Clerckx2017} \\ \hline
 
\multirow{3}{*}{MISO} & no CSIT		& UP 	& $\mathbf{w}_{n} = \frac{\sqrt{2P}}{\sqrt{NM}}$  & The uniform power allocation (UP) in MISO also simply assigns the same power to all frequencies and spacial components, with a zero phase. & \cite{Clerckx2016}  \\ \cline{2-6}

& CSIT		& UPMF	& $\mathbf{w}_{n} = \frac{\sqrt{2P}}{\sqrt{N}} \frac{\mathbf{h}_{n}^{H}}{\left\| \mathbf{h}_{n} \right\|}$   & The uniform power allocation (UP) is applied in the frequency domain, and the matched (or maximum ratio transmission) beamforming (MF) is applied in the spatial domain.     & \cite{Clerckx2016} \\ \cline{2-6}

& CSIT		& SMF	& $\mathbf{w}_{n} = \frac{\mathbf{h}_{n}^{H}}{\left\| \mathbf{h}_{n} \right\|} \left\| \mathbf{h}_{n} \right\|^{\beta} \sqrt{\frac{2P}{\sum_{n=0}^{N-1} \left\| \mathbf{h}_{n} \right\|^{2\beta}}} $& The single antenna channel gain $A_{n}$ and optimal phase $e^{-j\bar{\psi}_{n}}$ are substituted by the multi-antenna effective channel gain $\left\| \mathbf{h}_{n} \right\|$ and the MRT beamforming vector $\mathbf{h}_{n}^{H} / \left\| \mathbf{h}_{n} \right\|$, respectively.      & MISO version of \cite{Clerckx2017} \\ \hline

\bottomrule
\end{tabular}

\end{table*}
%
\par We consider conventional modulation schemes (commonly used and designed for communication purposes) such as PSK, QAM and Circularly Symmetric Complex Gaussian - CSCG (equally distributing power between the real and the imaginary dimensions, i.e., $\Re\left\{\omega\right\}\!\sim\!\mathcal{N}(0,P)$ and $\Im\left\{\omega\right\}\!\sim\!\mathcal{N}(0,P)$) and compare with modulations specifically designed for wireless power delivery, such as Real Gaussian (allocating the transmit power to only one dimension e.g. $\Re\left\{\omega\right\}\!\sim\!\mathcal{N}(0,2P)$) \cite{Varasteh2017}, and the recently proposed flash signaling \cite{Varasteh2018} characterized by a uniformly distributed phase $\phi$ over $\left[0,2\pi\right)$ and the amplitude distributed according to the following probability mass function
\par
\begin{align}\label{flash_distr}
p_s(s)=\begin{cases}
1-\frac{1}{l^2}, \ & s=0,\\
\frac{1}{l^2}, \ & s=l\sqrt{2P},
\end{cases}
\end{align}
with $l\geq 1$. We can easily verify that $\mathbb{E} \left[|\omega|^2\right]=\mathbb{E}\left[s^2\right]=2P$, hence satisfying the average power constraint. By increasing $l$, $s=l\sqrt{2P}$ increases and $p_s(l\sqrt{2P})$ decreases, therefore exhibiting a low probability of high amplitude signals.
%

\subsection{Transmit Diversity}

In contrast to (energy) waveform and modulation that induces amplitude fluctuations of the transmit signal, transmit diversity is designed to generate amplitude fluctuations of the wireless channel \cite{Clerckx2018}. Those fluctuations of the wireless channel are beneficial to the energy harvester thanks to the rectifier nonlinearity. 
To induce fluctuations of the wireless channel, transmit diversity relies in its simplest form on multiple dumb antennas fed with a low PAPR continuous wave and antenna-dependent time varying phases. In this case, the waveform design factor $\omega_{n,m}(t)$ at the antenna $m$ at time $t$ on carrier frequency $f_{0}$ is expressed as follows
\begin{equation}
\omega_{n,m}(t) = \omega_{m}(t) = s e^{j\phi_{m}(t)}, 
\end{equation}
where $s = \sqrt{\frac{2P}{M}}$ is the amplitude of the continuous wave on each transmit antenna (with uniform power allocation), and $\phi_{m}(t)$ is an antenna dependent time varying phase (whose rate of change can be predefined). The total transmit power over all antennas is fixed to $P$.
\par Transmit diversity can also be implemented in combination with the aforementioned energy modulation and waveform strategies.
Transmit diversity with energy modulation can be designed by transmitting the same energy symbol on all antennas but applying an additional antenna-dependent random phase $\phi_{m,\mathrm{td}}(t)$, such that 
\begin{equation}
\omega_{n,m}(t) = \omega_{m}(t) = s(t)e^{j\phi_{m}(t)},
\end{equation}
where $s(t) = \sqrt{\frac{2P}{M}}\sqrt{m_{I}(t)^{2} + m_{Q}(t)^{2}}$ and $\phi_{m}(t) = \tan^{-1} \Big ( \frac{m_{Q}(t)}{m_{I}(t)} \Big)+\phi_{m,\mathrm{td}}(t)$. 
The normalized complex modulation symbol $m(t)=m_{I}(t) + jm_{Q}(t)$ is the same for all antennas. Similarly, transmit diversity with multisine waveform transmits the same waveform on all antennas and applies an antenna dependent time varying phase before being launched over the air. Considering a channel non-adaptive in-phase multisine waveform with uniform power allocation in frequency and space (denoted as UP in Table \ref{wave_desc}), $\omega_{n,m}(t)$ on antenna $m$ at time $t$ on frequency $f_{n}$ is expressed as follows
\begin{equation}
\omega_{n,m}(t) = \omega_{n,m}(t) = s e^{j\phi_{m}(t)}, 
\end{equation}
where $s = \sqrt{\frac{2P}{NM}}$ and $\phi_{m}(t)$ is the antenna dependent time varying phase of transmit diversity.
%
%
\section{Theoretical Performance Analysis}\label{sec:waveform_sec}

The scaling laws of \eqref{diode_model_2} have been introduced in \cite{Clerckx2016} as a way to predict the theoretical performance benefits of WPT signal designs and the key role played by the rectifier nonlinearity and the signal parameters (e.g. $N$, $M$). The behavior predicted from the scaling laws will be contrasted with the measurement results. To that end, this section summarizes some of those existing theoretical scaling laws for waveform designs \cite{Clerckx2016} and for transmit diversity \cite{Clerckx2018}, and extends them to account for mobility conditions and to (energy) modulation.
%
%
%
\subsection{(Energy) Waveform and Beamforming}\label{subsec:wav_design}
%
\begin{table*}[h]
\centering
\caption{Scaling Laws of Energy Waveforms}
\label{scaling_laws}
\renewcommand{\arraystretch}{1.05} 

\begin{tabular}{>{\centering}m{20mm}>{\centering}m{25mm}>{\centering}m{40mm}>{\centering}m{70mm}}

\toprule 
\multirow{2}{*}{$z_{\mathrm{DC}}$}	&\multirow{2}{*}{N,M}	& No CSIT 											& CSIT						\tabularnewline
						&					& $z_{\mathrm{DC,UP}}$											& $z_{\mathrm{DC,UPMF}}$  		\tabularnewline \midrule \midrule
\multirow{2}{*}{FF Channel}	&N $\gg$ 1, M = 1		& $k_{2}R_\mathrm{ant}P + 2k_{4}R_\mathrm{ant}^{2}P^{2}N$	& $k_{2}R_\mathrm{ant}P + 2k_{4}R_\mathrm{ant}^{2}P^{2}N$		\tabularnewline \cline{2-4}
						&N $\gg$ 1, M $\gg$ 1	& $k_{2}R_\mathrm{ant}P + 2k_{4}R_\mathrm{ant}^{2}P^{2}N$	& $\epsilon^{2} k_{2}R_\mathrm{ant}PM + (1-\epsilon^{2})k_{2}R_\mathrm{ant}P + \epsilon^{4} k_{4}R_\mathrm{ant}^{2}P^{2}NM^{2} + 2(1-\epsilon^{2})^{2}k_{4}R_\mathrm{ant}^{2}P^{2}N$		\tabularnewline \midrule[1pt]
\multirow{2}{*}{FS Channel}	&N $\gg$ 1, M = 1		& $k_{2}R_\mathrm{ant}P + 3k_{4}R_\mathrm{ant}^{2}P^{2}$	& $k_{2}R_\mathrm{ant}P + 3k_{4}R_\mathrm{ant}^{2}P^{2} + \epsilon^{4}\pi^{2}/16 k_{4}R_\mathrm{ant}^{2}P^{2}N$			\tabularnewline  \cline{2-4}
						&N $\gg$ 1, M $\gg$ 1	& $k_{2}R_\mathrm{ant}P + 3k_{4}R_\mathrm{ant}^{2}P^{2}$	& $\epsilon^{2} k_{2}R_\mathrm{ant}PM + (1-\epsilon^{2})k_{2}R_\mathrm{ant}P + \epsilon^{4} k_{4}R_\mathrm{ant}^{2}P^{2}NM^{2} + 3(1-\epsilon^{2})^{2} k_{4}R_\mathrm{ant}^{2}P^{2}$			\tabularnewline \midrule \bottomrule
\end{tabular}

\end{table*}
%
\par
The scaling laws for waveform designs under perfect CSIT are provided in \cite{Clerckx2016}. We here extend them to account for delayed CSIT due to mobility and time varying channels. 
To represent the delayed CSIT in a mobility condition and account for the differences between the CSI acquired at the time of channel estimation and the actual channel at the time of transmission, we have added a channel instance factor $k$ to the transmit and receive signal.  
The evolution of $\mathbf{h}_{k,n}$ is modeled by a first-order Gauss-Markov process 
\begin{equation}
 \mathbf{h}_{k,n} = \epsilon\mathbf{h}_{k-1,n}+\sqrt{1-\epsilon^{2}}\mathbf{g}_{k,n},
\end{equation}
where $\mathbf{g}_{k,n} \in \mathbb{C}^{1\times M}$ has i.i.d. entries distributed according to $\mathcal{C}\mathcal{N}(0,1)$ and $\mathbb{E}\left[ \mathbf{h}_{k-1,n}^{*}\mathbf{g}_{k,n} \right]=\mathbf{0}_{M}$, where $\mathbf{0}_{M}$ denotes $M \times M$ zero matrix. 
We assume $\mathbf{g}_{k,n}$ is i.i.d for all frequency componets $n$ in FS channel, and $\mathbf{g}_{k,n}=\mathbf{g}_{k} \ \forall n$ in FF channel.
$\mathbf{h}_{0,n}$ is independent of $\mathbf{g}_{k,n}$ for all $k \geq 1$.
The coefficient $\epsilon(0 \leq \epsilon < 1)$ quantifies the amount of the correlation between elements $h_{k-1,n,m}$ and $h_{k,n,m}$, and we assume all the elements of $\mathbf{h}_{k,n}$ have the same $\epsilon$. 
The time correlation coefficient $\epsilon$ follows Jakes' model for fading channel \cite{proakis2008digital} $\epsilon = J_{0}(2 \pi f_{D}T)$ where $J_{0}(.)$ is the zeroth order Bessel function, $T$ denotes the channel instantiation interval, and $f_{D} = \frac{v f_{c}}{c}$ is the maximum Doppler frequency where $v$ is the terminal velocity, $f_{c}$ is carrier frequency, and $c=3 \times 10^{8} m/s$. The time correlation coefficient $\epsilon$ is therefore a measure of the channel time variation, and it is related to the velocity of the mobile terminal ($0 \leq \epsilon \leq 1 $).
\par
Following the same approach as in \cite{Clerckx2016}, we calculated the scaling laws of UP and UPMF techniques with the above delayed CSIT model in single and multi-antenna systems with frequency flat and selective channels. To that end, we assumed that the transmitter at time $k$ does not know $\mathbf{h}_{k,n}$, but has only access to the channel at time $k-1$ to design the transmit signal (i.e. a delayed version of the CSI). The scaling laws are shown in Table \ref{scaling_laws}.
\par Since the UP strategy is non-adaptive to the CSI, the time correlation coefficient $\epsilon$ does not affect its performance. The results of $z_{\mathrm{DC,UP}}$ in Table \ref{scaling_laws} is indeed not a function of $\epsilon$. A waveform gain proportional to $N$ is achieved in FF channels, but not in FS channels. No beamforming gain is achieved either. However, with the channel-adaptive UPMF strategy, $\epsilon$ has a significant effect on the $z_{\mathrm{DC}}$ performance. When $\epsilon=1$, the scaling laws $z_{\mathrm{DC,UPMF}}$ boil down to those provided in \cite{Clerckx2016}, and a gain proportional to $N$ and $M$ is observed in both FF and FS channel conditions. 
On the other hand, as $\epsilon$ decreases and approaches 0, $z_{\mathrm{DC,UPMF}}$ converges to $z_{\mathrm{DC,UP}}$. As $\epsilon$ decreases, the beamforming gain vanishes in FS and FF channels, while the waveform gain vanishes in FS channels but remains in FF channels. In other words, velocity and delayed CSIT incurs a bigger hit in FS channels than in FF channels.

%
\subsection{(Energy) Modulation} \label{modul_intro}

\par
This subsection derives the theoretical scaling laws of $z_{\mathrm{DC}}$ for each modulated signal. The transmission is assumed narrowband and the channel assumed frequency flat. We can write
\begin{align}
 z_{\mathrm{DC}} &= k_{2}R_\mathrm{ant}\mathbb{E}\{\left|m(t)\right|^{2}\}P + \frac{3}{2}k_{4}R_\mathrm{ant}^{2}\mathbb{E}\{\left|m(t)\right|^{4}\}P^{2}\nonumber \\
 &= k_{2}R_\mathrm{ant}P + \frac{3}{2}k_{4}R_\mathrm{ant}^{2}\mathbb{E}\{\left|m(t)\right|^{4}\}P^{2}
\end{align}
\vspace{-0.1cm}
where $m(t)$ is the normalized complex modulation symbol mentioned in section \ref{subsec:mod_design}. Since all modulations are normalized to have the same average transmit power, the difference between modulations can only be explained by the high-order moments, namely $\mathbb{E}\{\left|m(t)\right|^{4}\}$. Table \ref{table:mod_table} displayed $z_{\mathrm{DC}}$ of several modulation schemes such as PSK, QAM, Gaussians, and flash signaling and compare with the unmodulated Continuous Wave (CW).  
%
\begin{table}[!h]
\centering
\caption{Scaling Laws of Energy Modulation}
\label{table:mod_table}
\renewcommand{\arraystretch}{1.05}
\begin{tabular}{>{\centering}m{30mm} >{\centering}m{42mm}}
 \toprule
                   	 & $z_{\mathrm{DC}}$ 									 \tabularnewline  \midrule[1.5pt] 
Continuous Wave (CW)    & $k_{2}R_\mathrm{ant}P + 1.5k_{4}R_\mathrm{ant}^{2}P^{2}$          \tabularnewline \midrule
BPSK 		         	 & $k_{2}R_\mathrm{ant}P + 1.5k_{4}R_\mathrm{ant}^{2}P^{2}$			\tabularnewline	 \midrule
16QAM 		         	 & $k_{2}R_\mathrm{ant}P + 1.98k_{4}R_\mathrm{ant}^{2}P^{2}$		\tabularnewline	 \midrule
Complex Gaussian 		 & $k_{2}R_\mathrm{ant}P + 3k_{4}R_\mathrm{ant}^{2}P^{2}$		\tabularnewline	 \midrule
Real Gaussian 		 & $k_{2}R_\mathrm{ant}P + 4.5k_{4}R_\mathrm{ant}^{2}P^{2}$		\tabularnewline	 \midrule
Flash Signaling (with $l$) 		 & $k_{2}R_\mathrm{ant}P + \frac{3}{2}l^{2}k_{4}R_\mathrm{ant}^{2}P^{2}$		\tabularnewline	 \midrule
\bottomrule
\end{tabular}

\end{table}
\par
A first observation is that the second order term of $z_{\mathrm{DC}}$ and \eqref{diode_model_2}, i.e. the linear model of the EH \cite{Zeng2017,Clerckx2016}, is the same for all modulation schemes, cannot motivate the design of energy modulation and cannot predict the performance of energy modulation. 
A second observation is that there is a large performance gap between conventional modulations and those designed for WPT. This is due to the rectifier nonlinearity that favours modulations with large high- order moments $\mathbb{E}\{\left|m(t)\right|^{4}\}$. Among the conventional modulation methods, the complex gaussian (CSCG) signal shows the largest fourth order term compared to BPSK or 16QAM. A real Gaussian, though suboptimal for communication purposes, is more suitable for WPT since it leads to a higher fourth order moment than its complex counterpart. Flash signaling further boosts the fourth order term as $l$ increases. For $l>\sqrt{3}$, flash signaling is expected to lead to a higher DC power than a real Gaussian.
%
\subsection{Transmit Diversity}\label{sub:td_theory}
The performance of transmit diversity was analyzed in \cite{Clerckx2018}.
It was shown that by randomly changing the phase of a continuous wave on each transmit antenna, we achieve a gain proportional to the number of antennas $M$ in the fourth order term of $z_{\mathrm{DC}}$, despite the lack of CSIT. Additional benefits are obtained by combining transmit diversity with (energy) modulation and waveform.
The scaling laws of $z_{\mathrm{DC}}$ for transmit diversity with continuous wave and modulation/multisine waveform versus the single antenna continuous wave are displayed in Table \ref{table:td_table}.
%
\begin{table}[h!]
\centering
\caption{Scaling Laws of Transmit Diversity \cite{Clerckx2018}}
\label{table:td_table}
\renewcommand{\arraystretch}{1.05}
\begin{tabular}{>{\centering}m{12mm} >{\centering}m{41mm} >{\centering}m{22mm}}
 \toprule
                   	 & $z_{\mathrm{DC}}$ 					& Gain				 \tabularnewline  \midrule[1.5pt] 
CW				& $k_{2}R_\mathrm{ant}P + \frac{3}{2}k_{4}R_\mathrm{ant}^{2}P^{2}$  &        \tabularnewline \midrule
TD-CW			& $k_{2}R_\mathrm{ant}P + \frac{3}{2}k_{4}R_\mathrm{ant}^{2}P^{2}G_{\mbox{td}}$  & $G_{\mbox{td}}=1+\frac{M-1}{M}$           \tabularnewline      \midrule
TD-Modulation		& $k_{2}R_\mathrm{ant}P + \frac{3}{2}k_{4}R_\mathrm{ant}^{2}P^{2}G_{\mbox{td}}G_{\mbox{mod}}$  &$G_{\mbox{mod}}=\mathbb{E}\{\left|m(t)\right|^{4}\}$			\tabularnewline	 \midrule
TD-Multisine		& $k_{2}R_\mathrm{ant}P + \frac{3}{2}k_{4}R_\mathrm{ant}^{2}P^{2}G_{\mbox{td}}G_{\mbox{mt}}$  &$G_{\mbox{mt}}\stackrel{N\nearrow}{\approx}\frac{2N}{3}$		\tabularnewline	 \midrule

\bottomrule
\end{tabular}

\end{table}
%
%
\begin{figure*}[t]
	\centering
	\includegraphics[width=0.8\textwidth]{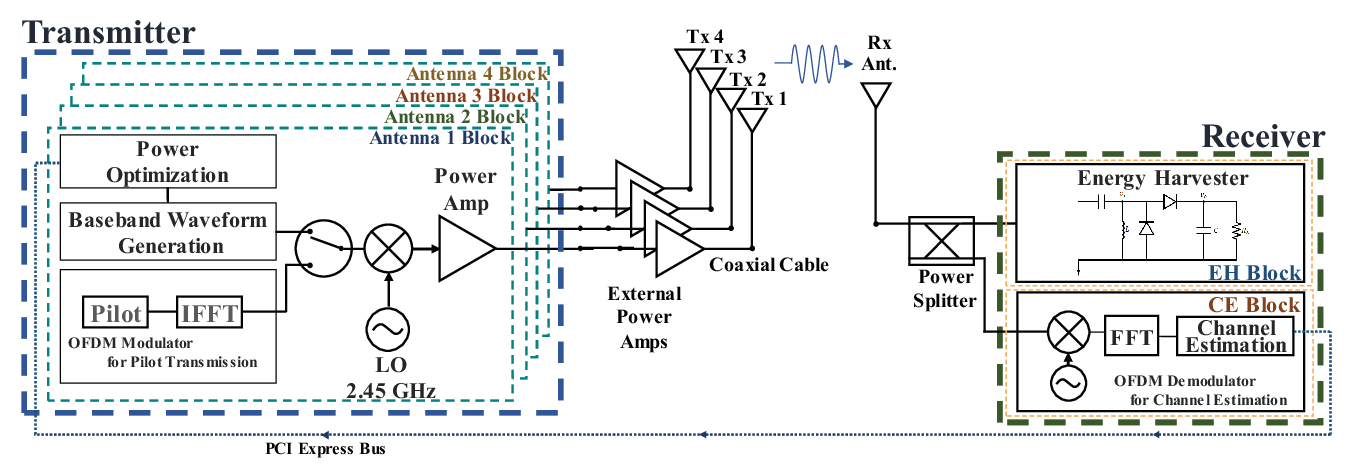}
	\caption{System architecture with equipment and a peripherals connection.}
	\label{sys_diag}
\end{figure*}

%
\section{Prototyping and Testbed Setup}\label{sec:prototype}
In order to verify that the proposed transmit signal design methods are feasible and improve the performance in a real world setting, a point-to-point WPT system prototype is required. This section discusses the implementation of a WPT system consisting of a transmitter capable of generating and transmitting various types of signals, and a receiver capable of channel estimation and energy harvesting. This system enables performance evaluation and validation of various signal generation techniques under various wireless channel environments\footnote{It can also be used to perform simultaneous wireless information and power transfer (SWIPT) in the future.}.

\subsection{Overall System Architecture and Hardware Setup}

\par
\begin{figure}[h]
	\centering
	\subfigure[]{\includegraphics[width=0.33\textwidth]{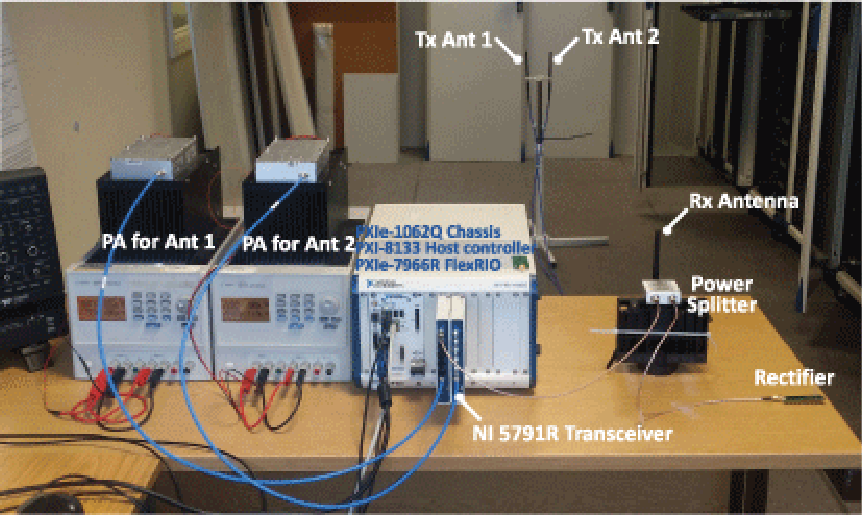}}
	\subfigure[]{\includegraphics[width=0.14\textwidth]{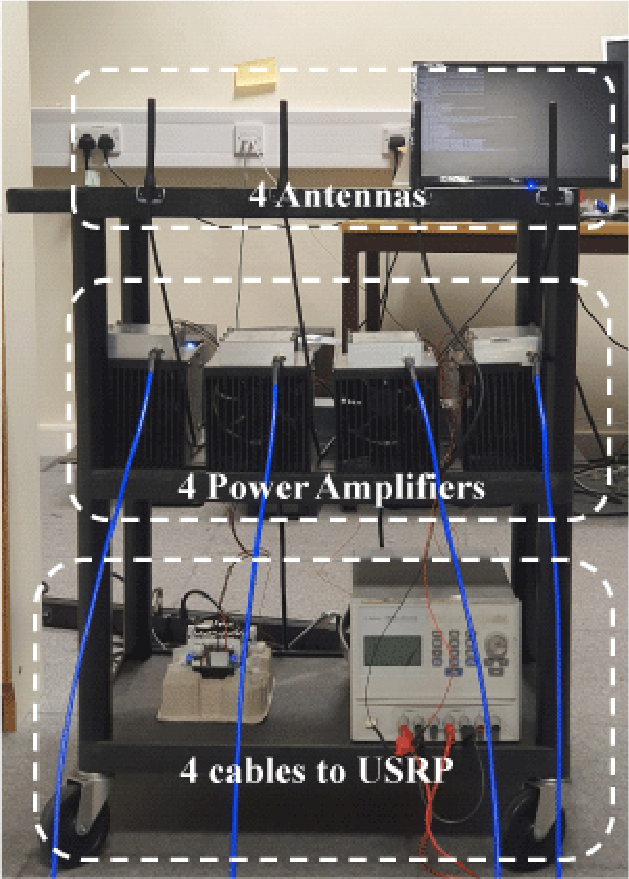}}
	\caption{WPT prototype (a) two antenna configuration (b) four antenna transmit antennas.}
	\label{equipment}
\end{figure}

 The system operates in the 2.4 GHz ISM band. The target operating range is to achieve an average received power of the order of -20 dBm at a distance of 5 meters. This is motivated by the fact that 10-100 $\mu W$ is enough to power modern wireless sensors and low-power devices \cite{Clerckx2018b}. In compliance with the Code of Federal Regulations, Title 47, Part15 (FCC Part15) regulation, the system is designed to operate with an effective isotropic radiated power (EIRP) of less than 4 watt (36dBm) \cite{FCC}. The system consists of up to four transmit antennas and one receive antenna and can be operated in MISO or SISO mode depending on the transmit signal strategy considered. Fig.\ref{sys_diag} displays the system block diagram which includes the equipment and the peripheral connections. Fig.\ref{equipment} illustrates the complete prototype.

\par We chose National Instrument (NI) software-defined radio prototyping equipment to implement the transmitter that is able to generate and transmit various types of WPT and pilot signals.
The transmitter hardware has been configured with a NI PXI platform and USRPs. Four pairs of RF transceivers were used to implement the four transmit antennas. 
The functions of signal design, optimization and generation on one hand and pilot transmission/channel acquisition on the other hand are combined within the transmitter, and these functions are programmed and controlled using LabView.
\par The receiver is divided into two large functional blocks. One is a channel estimation block (CE block) that receives the pilot signal, estimates the channel, and feeds back to the transmitter. The other is an energy harvesting block (EH block), made of a rectifier, that converts the received RF signal to DC power. 
The RF signal received by the antenna passes through the power splitter\footnote{We use a power splitter for measurement convenience, such as monitoring an RF input power to the energy harvester. An RF switch could have been used instead of the power splitter and may be a better choice to maximize the received power at the energy harvester. Unlike a power splitter that distributes power by 50\% to each block, it can send 100\% of power to the energy harvester during the wireless power transmission phase. However, the objective of this paper is to compare the energy harvesting performance of various signal design techniques. Therefore, using a power splitter does not affect the performance comparison, but makes the system easier to implement.} and delivers power to each block.
For the single antenna system, the CE block is also implemented on the NI SDR platform by using independent RF transceiver and FPGA module. For the multi-antenna system, one of the FPGA and RF transceiver modules operate as a transmitter and a receiver's CE block at the same time. The transmit and receive signal paths in the same module are operated completely independent and do not affect each other.
We installed the hardware (a pair of FPGA module and RF transceiver) responsible for the CE block in the same PCI express chassis as the transmitter.
This configuration enables CSI feedback from the receiver to the transmitter via the PCI express bus, which allows the transmitter to recognize the changes of CSI in real time\footnote{A final WPT system would require an over-the-air CSI feedback. We here use a wired (instead of wireless) feedback of the CSI as this experimental setup is sufficient to answer the main questions and objectives raised in the paper, namely to assess experimentally the advantages of closed-loop and open-loop systematic signal designs for WPT (including waveform, beamforming, modulation, transmit diversity), confirm theory from measurement, and validate the crucial role played by the rectifier nonlinearity. Replacing the wired feedback of the CSI by a wireless counterpart, and accordingly implementing and validating the design of optimized WPT signals (joint waveform and beamforming) under limited feedback, is an important issue that is left for future works.}. The cables connecting the equipment and the antenna are long enough so that various wireless channel environments can also be measured.

\subsection{Channel Estimation and WPT Signal Transmission}

\par
\begin{figure}[h]
	\centering
	\includegraphics[width=0.48\textwidth]{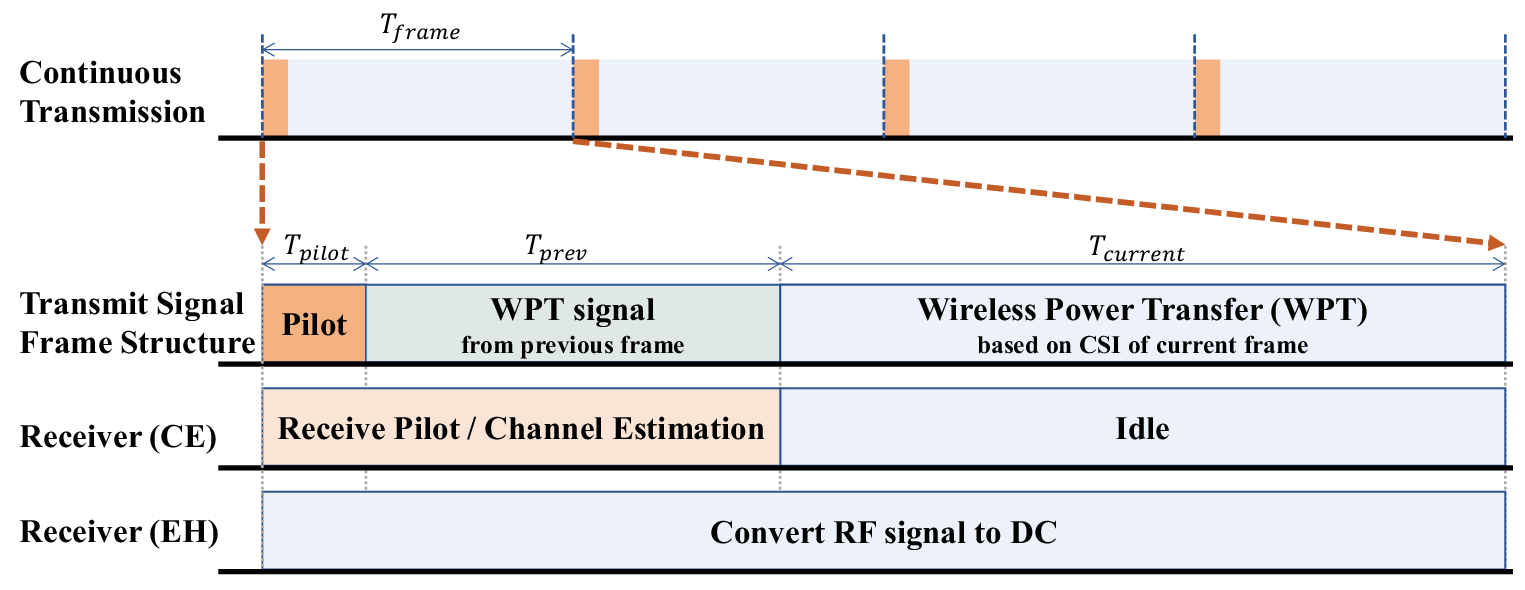}
	\caption{Frame structure and operations at the transmitter and receiver.}
	\label{frame}
\end{figure}

 
The architecture of Fig. \ref{sys_diag} requires the design of a suitable frame structure to enable channel acquisition and WPT signal transmission, as per Fig. \ref{frame}. The transmission signal includes two different types of signals, namely an OFDM signal for multi-frequency channel estimation and an optimized WPT signal (unmodulated multisine waveform or energy modulated continuous wave) for power delivery. The frame structure has therefore been designed to accommodate two different signals in the time domain. The length of the time frame $T_{frame}$ has been set by default to one second. One second was believed to be sufficient for deployments where the wireless channel does not change rapidly, such as in a static office environment and where there is no moving object during the measurements. Nevertheless, $T_{frame}$ can be adjusted and shortened to 200ms for deployments with moving objects. OFDM-based pilot signals are transmitted at the beginning of each frame for channel estimation and synchronization purposes. The duration $T_{pilot}$ has been fixed to 512 $\mu$s for single antenna transmission and includes therein a frame synchronization and pilot signals. 
In the case of multi-antenna transmission, the duration  $T_{pilot}$ is extended depending on the number of transmit antennas. To estimate multi-antenna channels, each antenna transmits a pilot in a different time slot. Therefore,  $T_{pilot}$ in the 2-antenna MISO experiment is 1 ms, and 4 ms for 4-antenna.
At the receiver, the CE block receives the pilot signal, estimates the channel, and feeds back the CSI to the transmitter. The transmitter then computes and generates an optimized WPT signal based on the calculated CSI. The time required for the computation and generation of the new signal (based on the newly acquired CSI) is $T_{prev}$ (approximately 30 to 40ms), and the signal optimized based on the CSI from the preceding frame is transmitted during this processing time. During the remainder of the frame, the wireless power signal optimized for the current frame (based on the current CSI) is transmitted and $T_{current}$ is usually within the range 960-970 ms.
\par The system uses a pilot-based channel estimation method. The pilot signal is generated based on OFDM signal for the estimation of the channel on a various number of frequencies. We use a block-type pilot that assigns a reference signal to all frequency components of interest. No interpolation is therefore needed. The Least-Square (LS) method is chosen as a channel estimation technique because of its low-complexity. The OFDM channel estimation block operates at 2.45 GHz center frequency with 20MHz bandwidth and subcarriers spacing of 78.125KHz. The upper and lower 5MHz bands are used as guard bands, thus the effective region that can actually be used to estimate the channel is the 10MHz in the middle and composed of 128 subcarriers. In other words, we can generate a maximum 128-tone signal and acquire the CSI on those 128 tones. The CSI is nevertheless commonly estimated on a smaller number of subcarriers, since the WPT optimized signal is transmitted on typically up to 16 tones because of the PAPR limits of the transmitter (that clips the signal when more than 16 in-phase sinewaves are transmitted). 

\par
\begin{figure}[!h]
	\centering
	\includegraphics[width=0.42\textwidth]{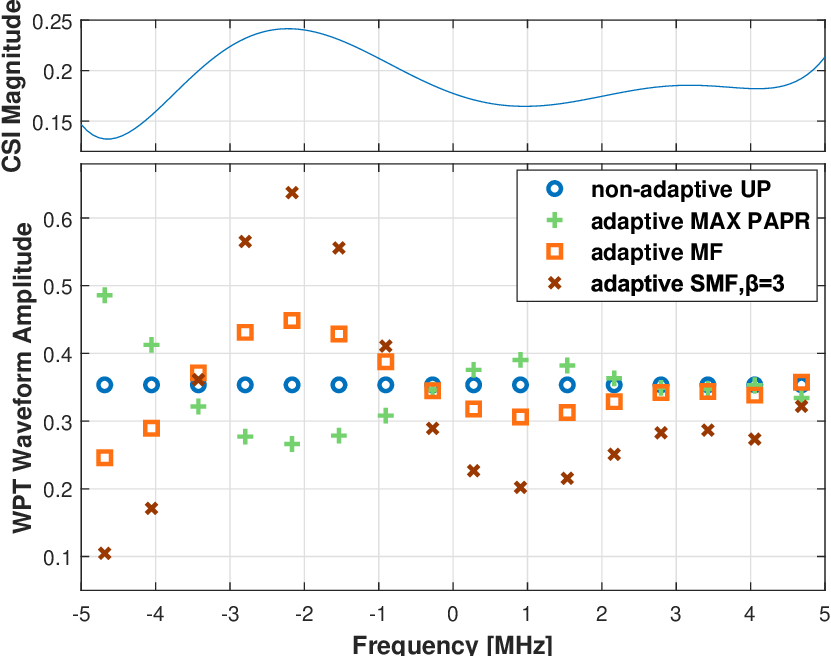}
	\caption{Frequency response (magnitude) of the wireless channel and WPT waveform magnitudes ($N=16$) for 10MHz bandwidth.}
	\label{power_alloc}
\end{figure}

WPT signals are generated based on the various signal design techniques introduced in Section \ref{sec:rect_model}. The channel adaptive multisine waveforms are applied to single and multi-antenna setups. The modulation signal is tested on a single antenna setup, and the transmit diversity signal is generated using two antennas. In order to illustrate the effect of the waveform designs of Table \ref{wave_desc}, Fig. \ref{power_alloc} displays the magnitude of a measured channel frequency response (for single antenna setup) and compares the allocated amplitudes for the different types of multisine waveform strategies. It can be seen that SMF allocates power to all frequencies (so as to exploit the rectifier nonlinearity), but emphasizes (more or less depending on the choice of $\beta$) the strong frequency components and attenuates the weakest ones (so as to benefit from the channel frequency diversity). This contrasts with MAX PAPR that inverts the channel (and allocates more power to the weakest components) so as to maximize the PAPR of the signal at the rectifier input. 

\begin{remark} Note that the proposed closed-loop architecture contrasts with conventional open-loop approaches in the RF literature with waveform being static/non-adaptive \cite{Trotter2009,Boaventura2011,Collado2014,Boaventura2013}, and beamforming relying on tags localization, not on the channel state \cite{Masotti2016}. Indeed, the waveform adaptation, channel estimation and frame structure are not present in those works, therefore preventing the signal at the input of the rectenna to be truly optimized. The proposed closed-loop architecture also differs from those of \cite{Choi2017,Choi2018,Yedavalli2017,Claessens2018,Abeywickrama2018} in the communication literature, where emphasis was on adaptive beamforming (to maximize $e_{\rf-\rf}$ ), rather than joint waveform and beamforming design (to maximize $e_{\rf-\rf}  \times e_{\rf-\dc}$).
\end{remark} 
\subsection{Rectifier Design}
\par To construct the receiver's EH block, we first considered a single-diode rectifier circuit. It consists of an impedance matching circuit, a diode and a smoothing circuit (low pass filter). The rectifier printed circuit board (PCB) is fabricated with a $\lambda$/4 length of microstrip, L-matching network, and followed by a Schottky diode rectifier circuit. The diode and the low pass filter implemented in the prototype are the same as in the rectenna used for circuit simulations in \cite{Clerckx2018}. The values of the matching network components have however been modified to fit the fabricated PCB and have been designed under the assumption of a 4-tone in-phase multisine input waveform as mentioned in \cite{Clerckx2017} under -20 dBm input power condition. The assumption of 4-tone input is chosen because it is a middle ground for all those test conditions (ranging from 1 tone to 16 tones). Also, though the input waveform can have 16 sinewaves, power allocation across all sinewaves is unlikely to be uniform due to the potential frequency selectivity. This implies that a subset of the sinewaves will be allocated power. Considering these cases, we have chosen the 4-tone as a robust baseline to design the rectifier. The reflection coefficient S11 of the rectifier is less than -10dB between 2.38 GHz and 2.5 GHz, and bandwidth is 120 MHz. We use Taoglas GW.15 antenna for the experiment. It is a universal 2.4 GHz band WiFi antenna, and the characteristics of the antenna are as follows: frequency 2.4-2.5 GHz, peak gain in free space <= 2dbi, efficiency <= 80\%, VSWR <=1.8. 
 
\par
\begin{figure}[h!]
	\centering
	\subfigure[]{\includegraphics[width=0.23\textwidth]{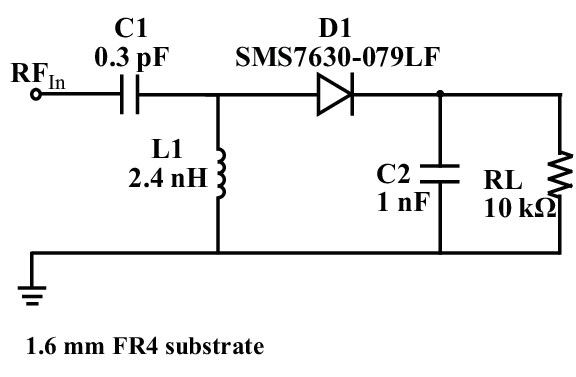}}
	\subfigure[]{\includegraphics[width=0.23\textwidth]{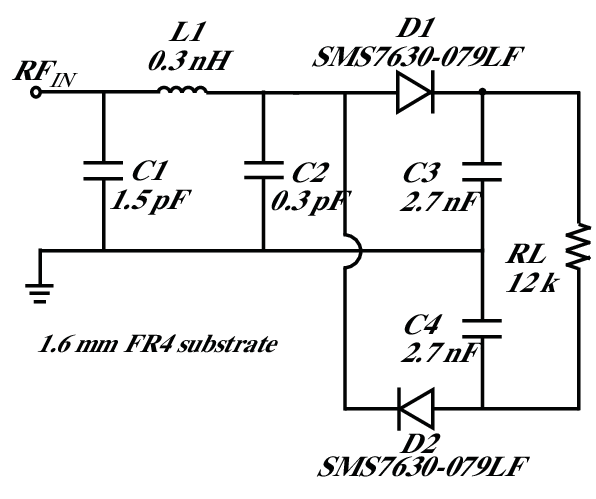}}\\
	\subfigure[]{\includegraphics[width=0.23\textwidth]{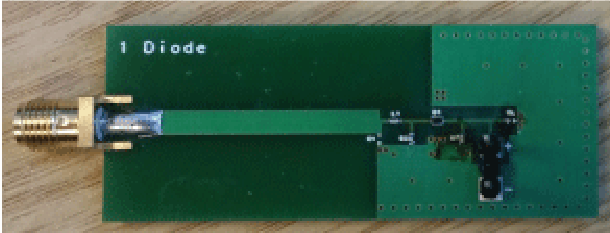}}
	\subfigure[]{\includegraphics[width=0.23\textwidth]{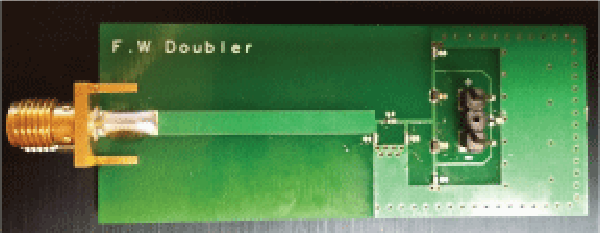}}
	\caption{Fabricated rectifiers and circuit schematics, Single-diode rectifier (a) schematic, (c) photo, and Voltage-doubler rectifier (b) schematic, (d) photo.}
	\label{rectifier_fab}
\end{figure}

 In addition, a rectifier with a voltage doubler structure was also built to verify the effectiveness of the nonlinear rectenna model and signal designs in other types of rectifier. The structure is the same as a single diode rectifier, but the output voltage is doubled using one rectifier for positive signals and one for negative signals, added via a series ouput. Circuit diagrams and photograph of the both rectifiers are shown in Fig. \ref{rectifier_fab}.


%

%
\section{Experiments and Validation}\label{sec:experiment}
The WPT testbed introduced in Section \ref{sec:prototype} has been experimented in various indoor propagation conditions. This section reports the measured harvested DC power for the various types of WPT signals. We compare the measured results with the observations made from the theoretical results of Section \ref{sec:waveform_sec}. We confirm experimentally the benefits of the systematic signal designs and the importance of modeling the rectifier nonlinearity in order to explain the measured results.
%

%
\subsection{Waveforms in SISO System}
The harvested DC power has been measured in various propagation environments with the objective to assess the impact of the multisine waveform design, the number of sinewaves and the bandwidth. Measurements were carried out in a normal office environment in static conditions. Test locations involve LoS and NLoS conditions, and exhibit frequency-flat (FF) channels and frequency-selective (FS) channels, respectively. 

\par The transmit waveforms are designed according to each waveform design schemes such as UP, MAX PAPR, ASS, MF, and SMF ($\beta$=3) with 1 to 16 tones of uniformly spaced sinewaves in 10MHz and 2.5MHz bandwidth. The inter-frequency spacing is given by $B/N$ with $B=10, 2.5$ MHz and $N=2,..,16$.  In all test cases, the transmit power was set to 33dBm and the RF power measured at the receiver based on the CW signal was around -20dBm. The single-diode rectifier of Fig. \ref{rectifier_fab}(a) was used.

\begin{figure}[h!]
	\centering
	\subfigure[FF channel, 10 MHz]{\includegraphics[width=0.39\textwidth]{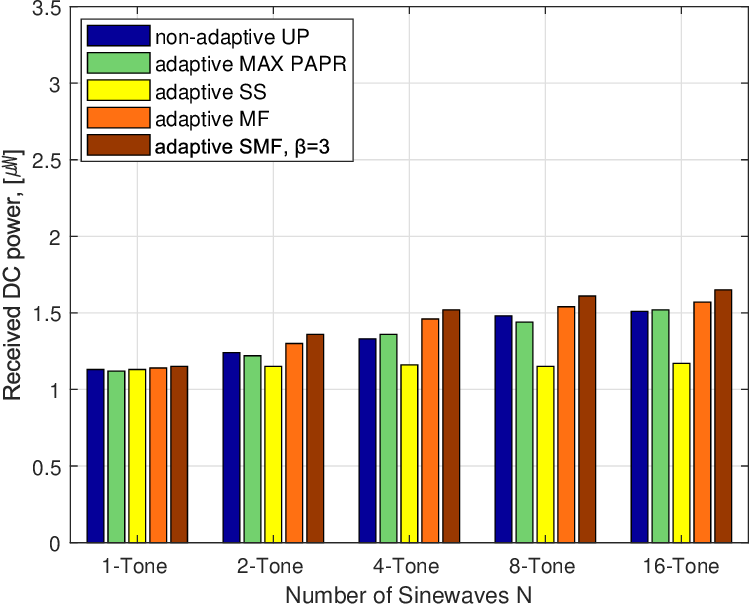}}
	\subfigure[FS channel, 10 MHz]{\includegraphics[width=0.39\textwidth]{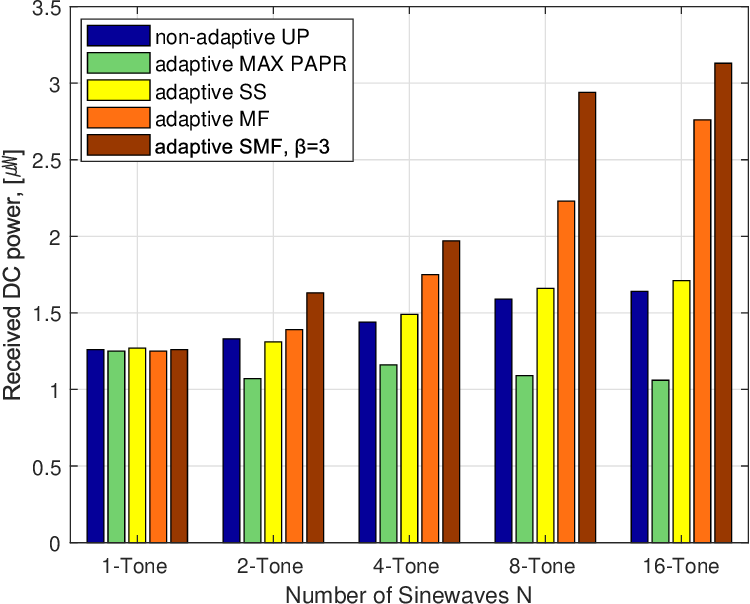}}
	\subfigure[FS channel, 2.5 MHz]{\includegraphics[width=0.39\textwidth]{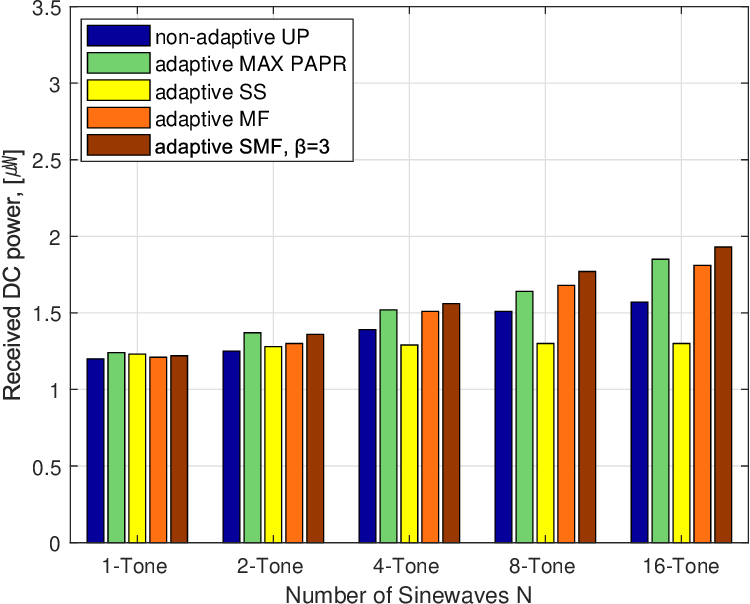}}
	\caption{Received DC power as a function of $N$ under various bandwidths and channel conditions: (a) Frequency flat channel and 10 MHz, (b) Frequency selective channel and 10 MHz, (c) Frequency selective channel and 2.5 MHz.}
	\label{siso_fffs_result}
\end{figure}

The harvested DC power was measured for 60 seconds for each test case and measurements were carried out five times\footnote{Splitting the total time duration into a number of short snapshots (60 seconds in this setup) is often used in channel characterization and measurement, e.g. \cite{Oestges2006}.}, with a 5min interval, at each location while maintaining static conditions, before taking the average.

\par Fig.\ref{siso_fffs_result} displays the received DC power measurement results as a function of $N$ under various bandwidths (2.5 and 10 MHz) and channel conditions (frequency flat and frequency selective). Since the test locations are different for FF and FS channel, the absolute value of the received power is different, but the relative performance gain according to different waveform design schemes in different channel characteristics can be observed. We make some important observations from the measurements.

\par \textit{First}, not all of the channel adaptive waveforms achieve better performance than the non-channel adaptive waveforms. The results of Adaptive SS (ASS) and MAX PAPR are indeed worse than UP in frequency-flat (FF) and frequency-selective (FS) channel, respectively. ASS allocates the full power to only one (though the strongest one) sinewave to maximize $e_{\rf-\rf}$, but at the cost of achieving a low $e_{\rf-\dc}$, and provides very little gain in FF channels because the waveform cannot benefit from any frequency diversity gain and does not exploit the rectifier nonlinearity. On the other hand, MAX PAPR scheme is inefficient in FS channel. MAX PAPR scheme inverts the channel to make the input waveform to the rectifier look like an in-phase multisine with uniform power allocation at the rectifier input. Therefore, MAX PAPR maximizes the PAPR of the input signal to maximize $e_{\rf-\dc}$ at the cost of wasting an excessive amount of power in inverting the channel and achieving a poor $e_{\rf-\rf}$. This confirms experimentally that focusing on maximizing PAPR in multisine waveform design (with the hope to maximize $e_{\rf-\dc}$), and allocating all power to the strongest sinewave (with the hope to maximize $e_{\rf-\rf}$) are not suitable strategies in general settings, as highlighted in \cite{Clerckx2016}.
\par \textit{Second}, increasing the number of sinewaves $N$ boosts the performance in FF and FS channels. By increasing $N$, a properly designed waveform can exploit the nonlinearity of the rectifier to boost $e_{\rf-\dc}$, but also exploits the frequency diversity of the channel to boost  $e_{\rf-\rf}$. This confirms results in \cite{Clerckx2016} that the diode nonlinearity is beneficial to WPT performance and is to be exploited in systematic waveform.  If $N$ increases continuously and the peak voltage increases above the breakdown voltage of the diode, the efficiency may decrease sharply. However, since $N$ is limited to 16 in the current prototype, the diode breakdown voltage is not reached. 
\par \textit{Third}, significant performance gain with a channel-adaptive waveform strategy such as SMF can be obtained in FS channel. Recall that an optimized waveform for WPT, including SMF, allocates power in a non-uniform manner to multiple sinewaves, with more power allocated to the strongest frequency components, so as to maximize $e_{\rf-\rf}\times e_{\rf-\dc}$ \cite{Zeng2017,Clerckx2016,Clerckx2017}. The gain of SMF with $\beta=3$ over non-adaptive UP with 16-tone on FF channel is 9.27\% but it reaches 90.85\% on the FS channel. Compared to conventional continuous wave (single tone), the gain is as high as 150\%. This confirms results in \cite{Clerckx2016} that CSI acquisition and systematic channel-adaptive waveforms that maximize $e_{\rf-\rf}  \times e_{\rf-\dc}$ are essential to boost the performance in frequency-selective channels (as in NLoS scenarios). 
In a SISO frequency-flat channel, the result also confirms that CSI is not essential to the transmitter to design efficient waveforms since there is no frequency selectivity to be exploited to further boost $e_{\rf-\rf}$.
\par \textit{Fourth}, comparing 2.5MHz and 10MHz bandwidth signals, we note that spreading the frequencies across a larger bandwidth is beneficial as the waveform design, if adaptive to the CSI, can benefit from a channel frequency diversity gain. This also confirms results in \cite{Clerckx2016} that larger bandwidths can boost the output DC power.
\par Overall, those observations are inline with the observations in the prior theoretical works \cite{Clerckx2016,Clerckx2017}, and with the theoretical gain of the waveform design that scales with $N$ (in the fourth-order term of \eqref{diode_model_2}) according to Table \ref{scaling_laws}. It is worth to recall that all those four observations were already made in \cite{Clerckx2016} and \cite{Clerckx2017} based on analysis and circuit simulations. 
All experimental results fully match with the theory and therefore validate the rectifier nonlinear model and the systematic waveform design methodology introduced in \cite{Clerckx2016}, \cite{Clerckx2017} and subsequent works \cite{Huang2017,Huang2018}. Results also confirm experimentally the feasibility and the promising gains offered by a closed-loop WPT architecture. 

\begin{remark} \label{remark_linear_model_waveform} The above results and observations also confirm experimentally the inaccuracy of the linear model, obtained by ignoring the fourth order term in \eqref{diode_model_2}, and its inefficiency in designing multisine waveforms \cite{Clerckx2016}. Recall that the ASS waveform is motivated by the linear model, as it results from allocating all power to the strongest frequency component \cite{Clerckx2016}. Clearly, the fact that the ASS performance is poor and even sometimes worse than non-adaptive waveforms demonstrate that the linear model does not capture the essence of the energy harvester, is inefficient for WPT signal designs, and is inaccurate to predict the waveform performance\footnote{ASS should achieve the highest performance according to the linear model, which is clearly not the case. Moreover the benefits of the other waveforms cannot be explained from the linear model.}. 
\end{remark}
%

%
\subsection{Waveforms with Voltage Doubler Rectifier}

In the previous subsection, we considered a rectifier composed of a single diode followed by a low-pass filter with a load $R_{\mathrm{L}}$, as illustrated in Fig. \ref{rectifier_fab}(a). This is the simplest rectifier configuration. In this subsection, the experiment is extended to other types of rectifiers with multiple diodes.

\par The nonlinear rectenna model was originally derived and motivated by a single diode rectifier circuit in \cite{Clerckx2016}. The model was then shown (analytically and through circuit simulations) to hold for more general rectifiers with many diodes in \cite{Clerckx2017}. In order to verify experimentally that the model and the corresponding signal designs are valid for other types of rectifier circuits with more diodes, the same test as in previous subsection has been performed using the voltage doubler circuit using two diodes of Fig. \ref{rectifier_fab}(b). 
\begin{figure}[h!]
	\centering
	\subfigure[FF channel]{\includegraphics[width=0.39\textwidth]{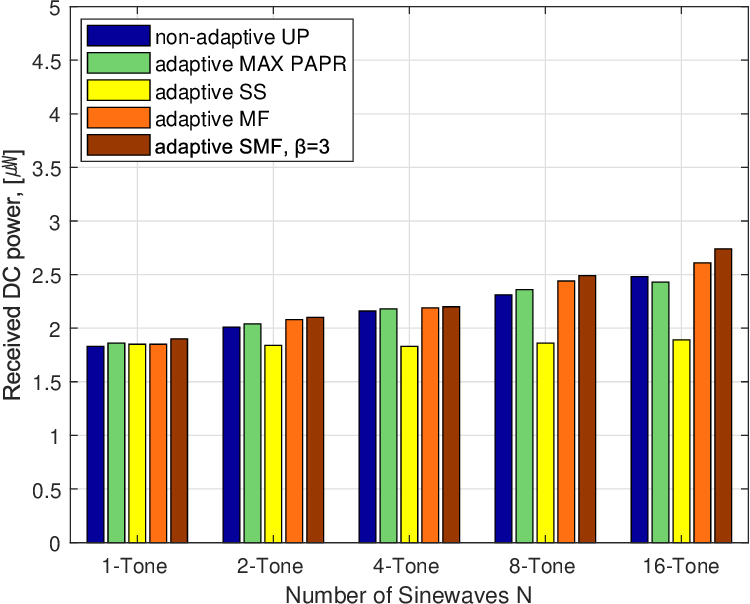}}
	\subfigure[FS channel]{\includegraphics[width=0.39\textwidth]{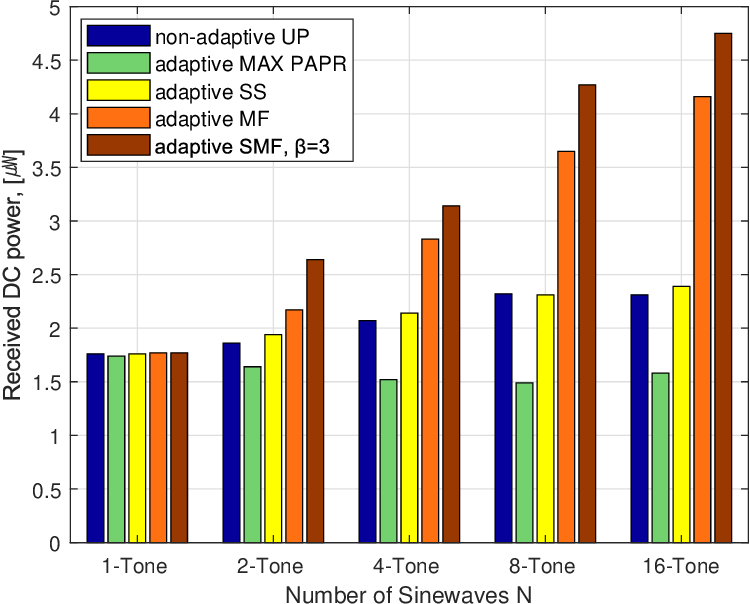}}
	\caption{Received DC power using voltage doubler rectifier as a function of $N$ with 10 MHz bandwidth under two different channel conditions (a) Frequency flat channel (b) Frequency selective channel.}
	\label{vd_fffs_result}
\end{figure}
\par It appears that the observations made from Fig. \ref{siso_fffs_result} with the single diode rectifier also hold for the voltage doubler rectifier in Fig. \ref{vd_fffs_result}. The increase and decrease trend of the harvested DC power as a function of the waveform designs remains the same for both rectifiers. The tests were performed in the same locations as the single diode rectifier experiment of Fig. \ref{siso_fffs_result}, and the overall received power increased by 30\% when using the voltage doubler. The SMF signal has a maximum gain over CW of 170\%, which is higher than that achieved in the single diode experiment. 
\par Results confirm that the nonlinear rectenna model \eqref{diode_model_2}, used for the design of systematic waveforms and for the prediction of the harvested DC power performance with various signal design techniques, is valid not only for a single diode rectifier circuit but also for a rectifier circuit using multiple diodes.
%

%
%
%
\subsection{Waveforms in Mobility Conditions}
WPT technology is expected to be predominantly embedded in low-power tiny and portable devices such as IoT devices. In the presence of mobility, CSI needs to be acquired on a regular basis. In the event where the channel changes rapidly between two successive CSI acquisition at the transmitter, the CSIT is delayed and the harvested DC power $z_{\mathrm{DC,UPMF}}$ drops due to a loss in waveform and beamforming gains, as shown in Section \ref{subsec:wav_design}. In this section, we investigate experimentally the sensitivity of channel-adaptive waveform to mobility.

\par We designed the experiment to check the relations between the channel state information acquisition period and the terminal velocity. In previous subsections, the time frame was fixed to one second, i.e. the CSI was acquired every second. In static channel conditions, such a time frame is sufficient but in mobility conditions, it may not be enough to guarantee a gain of channel-adaptive over non-adaptive waveforms. We here consider and compare two different frame structures, with 1 second and 200ms period, under various mobility profiles, with the objective to shed some light on the sensitivity of WPT signals to mobility. 
Different frame structures imply different channel acquisition periods. Since the period influences the time correlation coefficient $\epsilon$ mentioned in Section \ref{subsec:wav_design}, both frame structures experience different $\epsilon$ under the same velocity condition. 
We set four different velocity of moving antenna, namely 0.01, 0.05, 0.5, and 1 m/s, and investigate the gains over channel non-adaptive WPT. 1 m/s is approximately 4km/h which corresponds to pedestrian speed.

\begin{figure}[h!]
	\centering
	\includegraphics[width=0.4\textwidth]{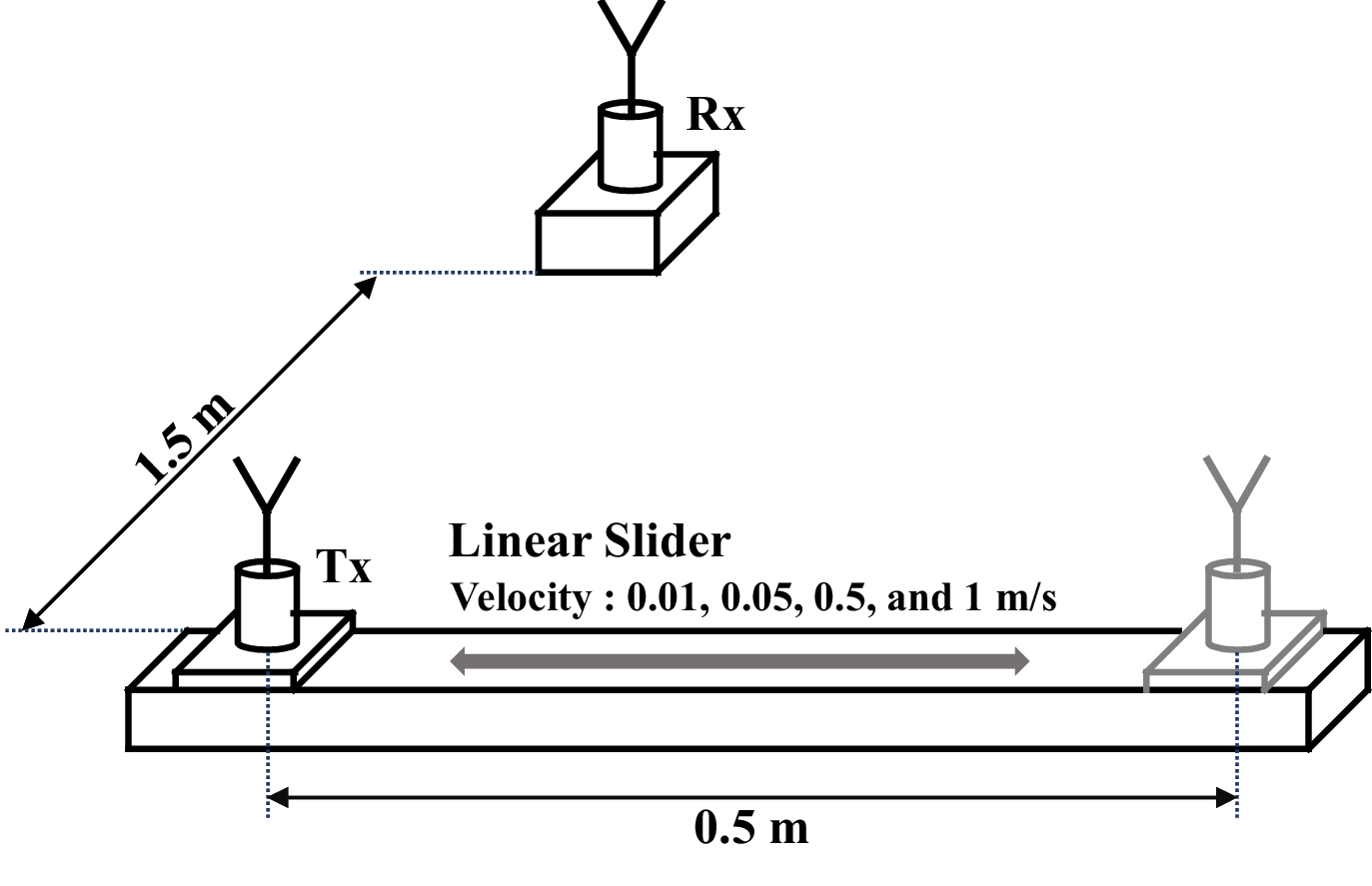}
	\caption{Mobility Experiment Setup.}
	\label{mobility_setup}
\end{figure}

\par To generate controllable and reproducible mobility conditions, we used a linear slider of 50cm length, illustrated in Fig. \ref{mobility_setup}, to move the transmitter while the receiver remains fixed. 
We compare the performance of a channel-adaptive SMF (with $\beta=3$) and a non-adaptive UP waveform, both consisting of 16 sinewaves uniformly spaced in a 10 MHz bandwidth. For each test case, measurements are carried out five times, each time taken for a duration of 5-minutes. Results are then averaged over all measurements. 
\begin{figure}[h!]
	\centering
	\subfigure[1s frame]{\includegraphics[width=0.39\textwidth]{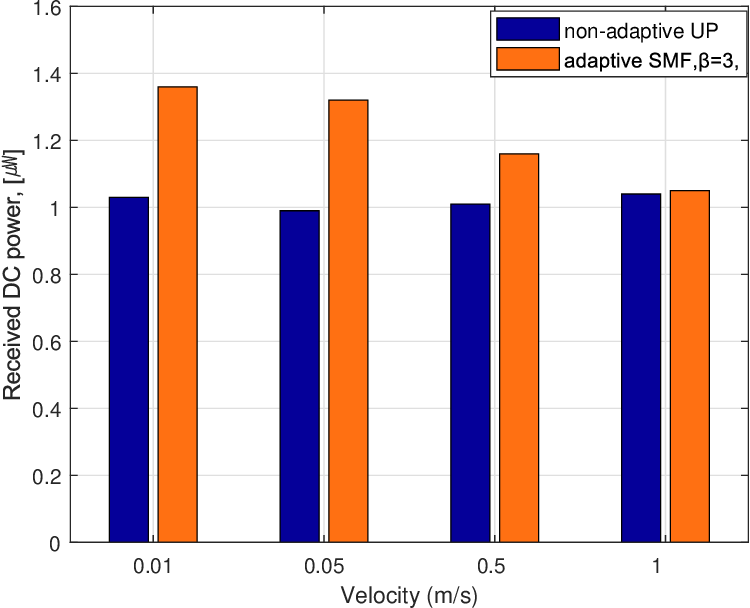}}
	\subfigure[200ms frame]{\includegraphics[width=0.39\textwidth]{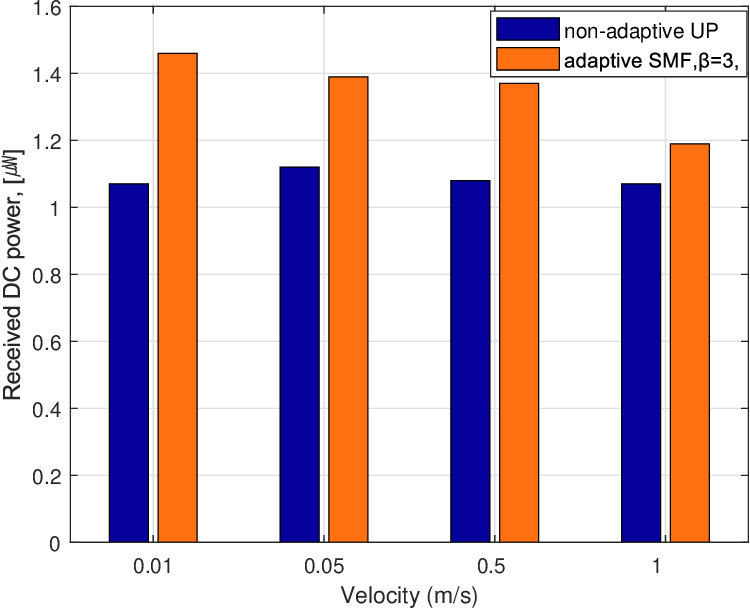}}
	\caption{Received DC power as a function of terminal velocity with different signal frame structures.}
	\label{mob_result}
\end{figure}
\par Fig. \ref{mob_result} shows the experimental results with the 1-second time frame and 200 ms time frame. 
The graph first shows that the harvested DC power level with the non-adaptive UP waveform is nearly constant regardless of the velocity of mobile antenna.
As shown in Section \ref{subsec:wav_design} the scaling law of the non-adaptive UP signal is not affected by the time correlation coefficient $\epsilon$, which is the same in the measurement results.
The graph also shows in both frame structure cases that the adaptive SMF signal exhibits some gain over non-adaptive UP signal in a low-velocity condition but the gain decreases as the velocity of the mobile antenna increases (i.e. as $\epsilon$ decreases).
Since $\epsilon$ is related not only to the velocity but also to the channel estimation period, the gain reduction rate of the SMF signal due to the increase in velocity is different in the two frame structures.
In a low-velocity test case such as 0.01m/s, adaptive SMF signals have a similar gain of about 40\% over non-adaptive UP for both 1s and 200ms frame structure. On the other hand, at 1m/s pedestrian velocity, the gain of the SMF signal is almost zero when using the 1s frame, but a gain of 12\% is still observed when using the 200ms frame.
\par 
These observations show the relation between the velocity of a mobile antenna, CSIT acquisition period, the time correlation coefficient $\epsilon$ , and the harvested DC power. 
The influence of $\epsilon$ on DC power harvesting performance shown in Section \ref{subsec:wav_design} was confirmed in this experiment.
The design of an appropriate frame structure is important to cope with various mobility conditions.

%

%
\subsection{Joint Beamforming and Waveform in MISO System}
The prototype system is equipped with two antennas, and performance can therefore benefit from a beamforming gain on top of the waveform gain already highlighted in previous subsections. According to the scaling laws in Table \ref{scaling_laws}, the beamforming and waveform gains are cumulative as both appear in the fourth order term of \eqref{diode_model_2} through the term $N M^2$. As discussed in \cite{Clerckx2016}, this highlights that the number of transmit antennas and number of sinewaves can be traded off to achieve a given target performance. 
In this subsection, we assess the performance benefits of conducting a joint beamforming and waveform design over the single-antenna waveform design and over conventional multi-antenna energy beamforming with continuous wave \cite{Zeng2017}.


%
In other words, we assess the performance benefits of exploiting jointly the spatial (beamforming) and frequency (waveform) domains of the transmit signal, and investigate how one could leverage the frequency domain of the waveform to decrease the complexity of the spatial domain beamformer (number of transmit antennas).

\begin{figure}[h!]
	\centering
	\subfigure[FF channel]{\includegraphics[width=0.39\textwidth]{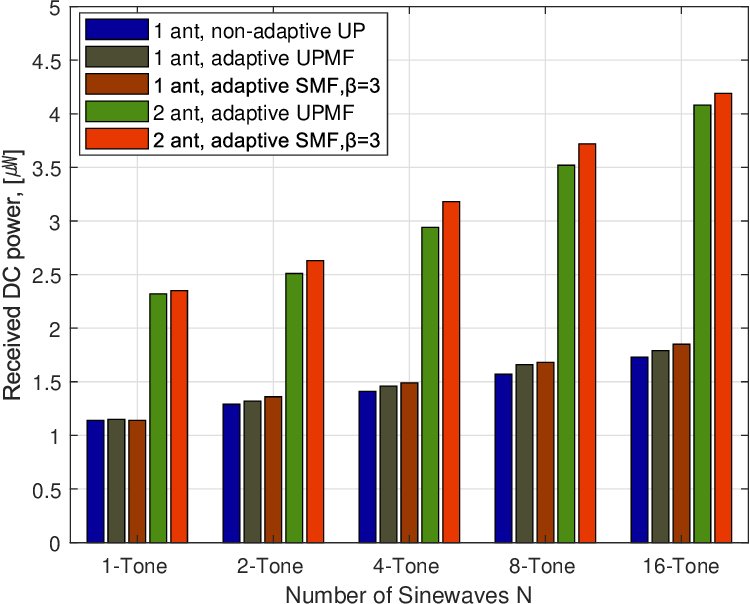}}
	\subfigure[FS channel]{\includegraphics[width=0.39\textwidth]{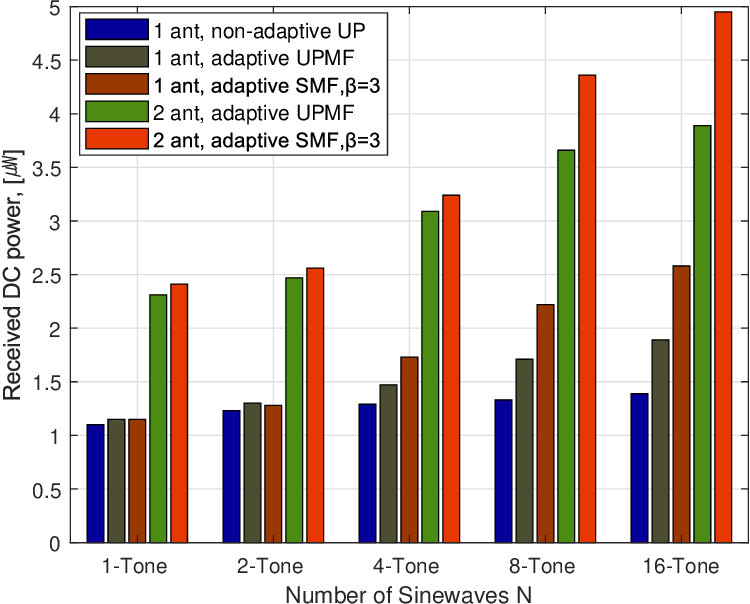}}
	\subfigure[CDF]{\includegraphics[width=0.39\textwidth]{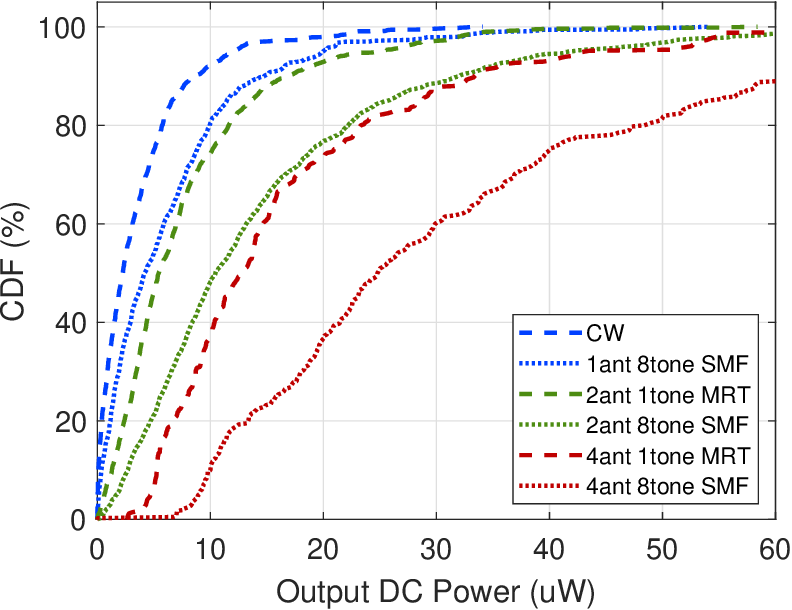}\label{cdf}}
	\caption{(a), (b)Received DC power as a function of $N$ with 10 MHz bandwidth under two different channel conditions (FF and FS) with one and two transmit antennas. (c) Measured CDF of DC output power in different locations}
	\label{miso_fffs_result}
\end{figure}

\par The experiments were performed in FF (LoS position) and FS (NLoS position) channel conditions as in the single antenna system. UPMF and SMF signals on two antennas and UP, UPMF, and SMF signal on one antenna are used for performance comparison for various $N$. Recall that UPMF in single-antenna setting relies on CSIT for channel phase compensation on each sub-carrier (in contrast to UP) and allocates power uniformly over all sub-carriers (similarly to UP). The experiments were carried out at five different locations for each FF and FS channel condition. Test locations were chosen to have FF channel on LoS position and FS channel on NLoS position with received RF power of about -20dBm based on CW signal. The harvested DC power was measured for 60 seconds, repeated five times, and results were averaged over all measurement for each test case. Fig. \ref{miso_fffs_result} displays the harvested DC power for each signal design and number of tones. We make the following observations.
\par \textit{First}, we observe that spatial domain and frequency domain processing can be traded off. Comparing 2-antenna SMF with CW ($N=1$) and 1-antenna SMF with $N\geq 16$, we note that the 1-antenna waveform outperforms the 2-antenna beamforming in FS channel. This significant gain can be obtained in FS channel where the 1-antenna SMF with $N\geq 16$ can jointly exploit the nonlinearity of the rectifier and the channel frequency diversity. This shows that the hardware complexity increase in the spatial domain (having two antennas rather than one) can be decreased by adopting a more efficient waveform. In other words, the use of SMF multisine waveform can decrease the need for multiple transmit antennas for a given performance target. 

\par \textit{Second}, we observe that the gains from the spatial (beamforming) and frequency (waveform) domains are cumulative. For UPMF and SMF, the 2-antenna setting leads to about 100\% gain over the SISO setting for all $N$ in both channels. Remarkably, the 2-antenna SMF with $N=16$ achieves a gain of about 110\% over the 2-antenna conventional beamforming with CW ($N=1$) and close to 350\% over the 1-antenna CW, in FS channel conditions. Interestingly, the sharp increase in DC power with $N$ achieved by the 1-antenna SMF is still observed in the 2-antenna setting. This highlights that SMF jointly exploits the multi-antenna beamforming gain, the channel frequency diversity gain and the rectenna nonlinearity. Those results also show that various performance enhancement factors can be superimposed and applied in WPT, which can lead to significant performance improvements.
\par
Additional MISO joint beamforming and waveform experiments were carried out using the four antenna prototype to verify the performance of multi-antenna with multi-tone WPT under many different wireless environments. 
Cumulative distribution function (CDF) of measured DC output power with various numbers of antennas and tones is presented in Fig. \ref{cdf}.
One tone MRT and eight tone SMF signals for each number of antennas were used and performance compared. 
The measurement are taken at 100 different locations in the office and the distance between the transmitter and the receiver varies between 3 and 5.5 meters. 
Each measurement was taken 1 second and repeated ten times with one minute of time interval at each location. 
The graph clearly shows multi-antenna with multi-tone SMF signals outperform multi-antenna with single-tone MRT signals, and the gain is significant.
Besides, the 4-antenna 1-tone waveform shows similar performance that of the 2-antenna 8-tone waveform. 
In the same manner, 2-antenna single-tone and 1-antenna 8-tone show similar performance. 
Such behavior reaffirms the two observations we identified earlier regarding the joint beamforming and waveform gain.
Those observations are inline with the observations made from the theoretical gain of the joint waveform and beamforming design that scales with $N M^2$ (in the fourth order term of \eqref{diode_model_2}) according to Table \ref{scaling_laws}. This indicates that the theoretical analysis and simulation results provided in Section \ref{subsec:wav_design} are consistent with the experimental results in the actual wireless environment. 
\begin{remark} 
The key observation of this subsection is that different types of gains can be accumulated by jointly using waveform and beamforming, such as a beamforming gain, a frequency diversity gain and the gain from the rectifier nonlinearity. 
This contrasts with beamforming-only approaches, e.g., \cite{Choi2017,Choi2018,Yedavalli2017}, that provide a beamforming gain-only. Recall that results of the beamforming-only approach (and therefore somewhat equivalent to \cite{Choi2017,Choi2018,Yedavalli2017}) is obtained by looking at 1-tone results in Fig. \ref{miso_fffs_result}. 
Results here show that the gain of the joint beamforming and waveform design scheme leads to significantly larger harvested DC power compared to the conventional beamforming-only schemes of \cite{Choi2017,Choi2018,Yedavalli2017}. .

\end{remark}
%
%
\subsection{Modulations}

\begin{figure}[h!]
	\centering
	\includegraphics[width=0.39\textwidth]{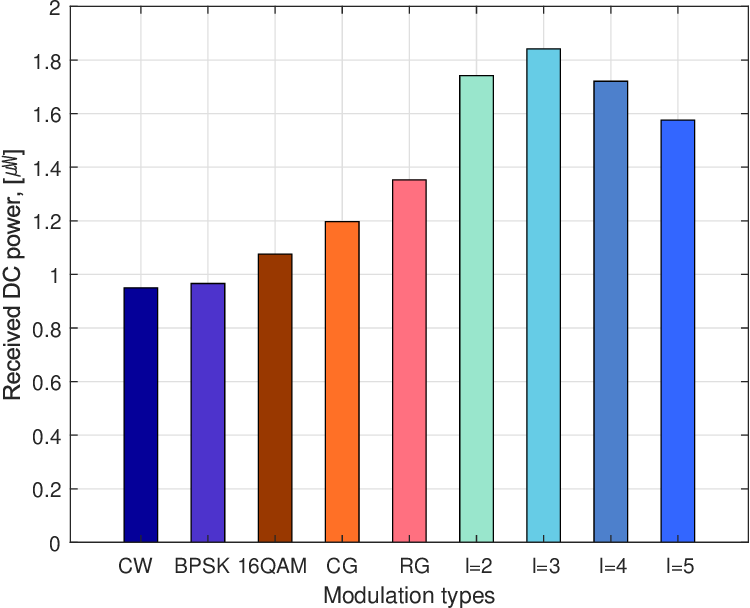}
	\caption{Received DC power vs. Modulation types.}
	\label{modul_exp_result}
\end{figure}

According to the scaling laws of Table \ref{table:mod_table}, conventional modulations used for communications such as BPSK, QAM and complex Gaussian (simply denoted as CG) should be outperformed by energy modulations such as real Gaussian (RG) and flash signaling. We carried out a modulated waveform experiment in order to confirm the theoretical predictions of Table \ref{table:mod_table}. The signal was generated with a modulation rate of 2.5 MHz for all modulation types. To rigorously observe the differences due to the modulation schemes (rather than measuring the effect induced by the fluctuations of the channel), the experiment was conducted by feeding the transmitted signal directly into the rectifier through cable connections (in contrast to the over-the-air radiation used in the other subsections). The rectenna received input RF power was set at -20dBm, and the harvested DC power was measured for five minutes and five times for each modulation type, before being averaged. Fig. \ref{modul_exp_result} displays the measurement results. 
\par We observe that the general trend of Fig. \ref{modul_exp_result} matches well the theoretical results of Table \ref{table:mod_table}. Namely, the PSK modulations does not perform any better than CW because PSK does not induce any amplitude fluctuation and does not affect the fourth order term $\mathbb{E}\{y(t)^{4}\}$ of $z_{\mathrm{DC}}$. 16QAM and CG exhibit a respective 13\% and 26\% gain compared to CW because of the amplitude fluctuation that increases $\mathbb{E}\{y(t)^{4}\}$. Similarly, RG achieves a 42\% gain thanks to the larger fourth moment of a RG distribution compared to a CG distribution. Flash signaling provides even higher DC power as it increases the fourth moment $\mathbb{E}\{y(t)^{4}\}$ as $l$ increases by enabling a small probability of very large amplitude signals. Nevertheless, the behavior does not match exactly what was predicted from Table \ref{table:mod_table}. The highest DC power is achieved at $l=3$ with an overall gain of 95\% over CW, but decreases when $l$ is further increased. This behavior is due to the peak voltage of the received signal that exceeds the breakdown voltage of the diode. Such breakdown voltage is not modeled in $z_{\mathrm{DC}}$. A similar trend, though less severe, has been observed in the circuit simulations provided in Appendix \ref{circuit_simulations}, though the breakdown voltage of the diode was found to be lower in the actual circuit than in the circuit simulations. Note that the performance could be improved by designing a circuit that is robust to diode breakdown and copes with high peak voltages (see discussions in \cite{Clerckx2018a,Clerckx2018b} and references therein). 

\begin{remark} It is important to recall that observations made from Fig. \ref{modul_exp_result} cannot be explained from the linear model of the rectenna. All those modulations achieve the same second order term $\mathbb{E}\{y(t)^{2}\}$, and according to the linear model, they should all achieve the same performance. Obviously this is incorrect and only accounting for the rectifier nonlinearity through the fourth order term $\mathbb{E}\{y(t)^{4}\}$ can explain the difference between the modulations. This further demonstrates that the inaccuracy of the linear model highlighted in Remark \ref{remark_linear_model_waveform} carries on to other types of signals such as modulation.  
\end{remark}

\begin{remark} 
Energy modulation is not only important for improving WPT efficiency but plays a major role in simultaneous wireless information and power transmission (SWIPT) systems \cite{Varasteh2017,Varasteh2018}. 
Conventional PSK/QAM modulations have been used in SWIPT \cite{Sakaki2014, Claessens2017}, and measurement results here show and confirm that we should depart from such modulation if one wants to make the best use of the radiowaves. 
Results show here that energy modulations based on flash signaling, whose randomness has been optimized to maximize the amount of DC power with no consideration for information transfer, significantly outperform other modulations/distribution in terms of harvested DC power.
It is left as future work to experimentally show how such modulations perform in terms of information transfer and what is the tradeoff between rate and power.   

\end{remark}
%

%
\subsection{Transmit Diversity}
\begin{figure}[h!]
	\centering
	\includegraphics[width=0.39\textwidth]{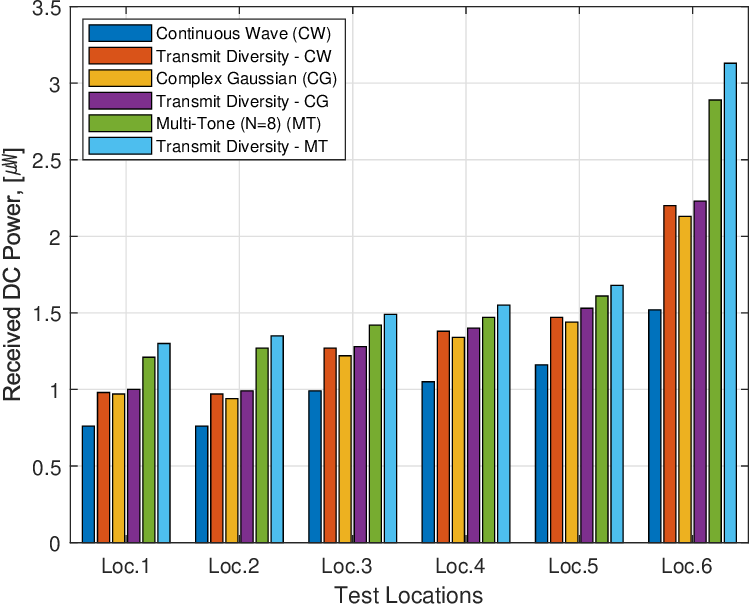}
	\caption{Transmit Diversity performance measurement in six different locations}
	\label{td_exp_result}
\end{figure}

The transmit diversity experiment was performed using two transmit antennas at six different LoS test locations located at a distance of 2.5 to 4m in a normal office environment. At each test location, the transmitter generates different types of signals such as single antenna continuous wave/complex Gaussian/multisine waveform ($N=8$) and their two-antenna transmit diversity counterparts. The phase changing rate for transmit diversity signals and modulation rate for the complex Gaussian signal is set to 2.5MHz. The DC power measurements were conducted for one minute and repeated five times with some time intervals, before being averaged to obtain the final measurements. Fig. \ref{td_exp_result} displays the measurement results at the six different test locations.

\par The experimental results show that the TD with CW signal has an average gain of about 28\% compared to the CW signal although there is some difference in each test position. TD with CG and TD with multisine/multitone ($N=8$) signal show a 31\% and a 66\% gain respectively over the CW signal. 
Those results are inline with the observations from the theoretical analysis of TD signals provided in Table \ref{table:td_table}.
This indicates that the theoretical model that shows the energy harvesting performance improvement by using TD in subsection \ref{sub:td_theory} is consistent with the actual experimental results over real-world wireless channels. Recall that those gains are achieved without any CSIT. Transmit diversity is appealing for low-complexity applications with a massive number of low-power devices because the transmitter is equipped with cheap/dumb antennas, the receivers do not need power-consuming signal processing block such as channel estimation and feedback, and the energy harvesting performance can be improved simultaneously for all receivers. Though the prototype was designed and measurements were conducted with two transmit antennas, as mentioned in the theoretical model, the gain can be improved by increasing the number of transmit antennas.
%

%
\section{Conclusions and Future Works}\label{sec:conclusion}
A WPT testbed with and without CSIT acquisition and various signal transmission strategies (beamforming, waveform, modulation and transmit diversity) was designed, prototyped and experimented. The harvested DC power achieved by those strategies and combination thereof was analyzed as a function of various parameters such as the propagation conditions, CSIT quality, bandwidth, rectenna design and experimental results were contrasted with the theoretical analysis.
\par It has been shown that the design of an appropriate signal generation method (such as SMF) that adapts as a function of the channel condition can significantly boost the harvested DC power performance. Large gains are obtained when using a combination of waveform and beamforming. The larger the number of tones in the waveform and the wider the bandwidth, the larger the gains. Significant performance improvements were possible through signal design based on CSIT under frequency-selective channel, so as to jointly benefit from a beamforming gain, a waveform gain, the rectenna nonlinearity and the frequency selectivity of the channel.
In the case where CSIT is not available, the power transmission efficiency can be greatly improved by using proper energy modulations or by generating artificial fading through a multi-antenna transmit diversity strategy. Widely used modulations for data communication have also been shown to improve the power transfer efficiency depending on the modulation type, but are outperformed by modulation designed specifically for WPT. 
This work demonstrates experimentally the importance and benefits of modelling and exploiting the harvester nonlinearities originating from the convexity of the I-V characteristics of the diodes. On the other hand, it is also verified that the linear model of the harvester obtained by ignoring the nonlinearity leads to poor signal design.
\par There are many interesting research avenues to pursue. Beyond the MISO system, a large-scale multisine multiantenna WPT with jointly optimized beamforming and waveform, applicable to both single-user and multi-user deployments, is a promising architecture \cite{Huang2017}. It is also worth to implement and investigate a larger number of transmit antennas in the transmit diversity experiment. When it comes to channel acquisition, the wired feedback of CSI needs to be replaced by a wireless counterpart. To that end, a low-power simple method to feedback CSI from receiver to transmitter and accordingly design the joint waveform and beamforming, as studied in \cite{Huang2018}, would be an interesting avenue that has not been experimented yet. 
Moreover, another interesting area will be to consider a WPT architecture where the transmit signals and the rectennas adapt themselves dynamically as a function of the channel state \cite{Clerckx2018b}, which requires the design of rectennas adaptive to their input waveforms (shape and power) \cite{Ouda2018}. These will be considered in future enhancements of our testbed system. Moving beyond WPT, it is also interesting to study how the prototype could be expanded to a real SWIPT system so as to assess the performance of SWIPT waveform and the corresponding rate-energy tradeoff. Some preliminary results are available in \cite{Kim2019}.
%

%

\section{Acknowledgments}
We thank B. Lavasani and National Instruments for providing some of the equipments needed to conduct the experiment.
%

%

\appendices
\section{Circuit Simulations}\label{circuit_simulations}

\par
\begin{figure}[h!]
\centering
\includegraphics[width=0.39\textwidth]{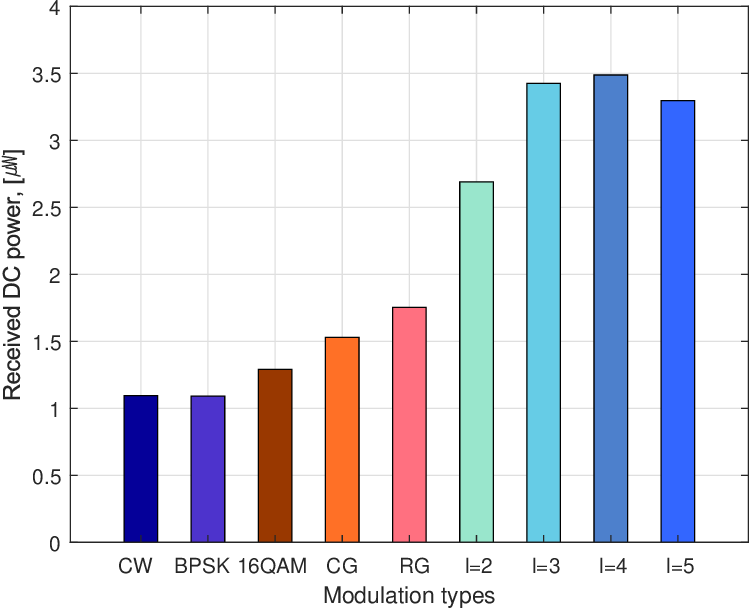}
\caption{Simulated received DC power with several modulation schemes and flash signaling $l=2,3,4,5$.}
\label{mod_simul}
\end{figure}

 Beamforming, waveform, modulation and transmit diversity performance have been analyzed using circuit simulations and results have been contrasted with the theory (using $z_{\mathrm{DC}}$ scaling laws). Readers are referred to \cite{Clerckx2016} and \cite{Clerckx2017} for waveform and beamforming, to \cite{Varasteh2018} for modulation and to \cite{Clerckx2018} for transmit diversity. In all cases, circuit simulations confirm the benefits of the four signal strategies.
In the sequel, we provide some more PSPICE circuit simulations for modulation to complement the ones obtained in \cite{Varasteh2018}.
The rectifier circuit for the simulation is the same as the one used in \cite{Clerckx2018} and we generate modulation signals with 2.5MHz symbol rate and -20dBm of RF Power in Matlab. The simulations were repeated 300 times using randomly generated modulation signals for each modulation format, and the results were then averaged.
Fig. \ref{mod_simul} illustrates the received DC power simulation results of different modulations.

\par The results show that some of the conventional modulations are effective to boost the DC power. For instance, CG signals exhibit higher efficiency than other conventional modulation schemes. PSK modulation has no performance advantage compared to the continuous wave because all symbols have the same magnitude. 16QAM signal leads to a performance improvement of about 17\% because of the amplitude fluctuations among symbols. On the other hand, with energy modulation, the performance improvement is more significant. RG leads to a 60\% gain compared to a continuous wave. Flash signaling exhibits significantly better performance. The maximum delivered power occurs at $l=4$. The gain observed on circuit simulations with flash signaling also appear larger than in the measurements of Fig. \ref{modul_exp_result}. 
The results also show that the simulations are inline with the scaling laws calculated in Table \ref{table:mod_table}.
%

\ifCLASSOPTIONcaptionsoff
  \newpage
\fi

\bibliographystyle{IEEEtran}
\bibliography{jhlib}

\end{document}